\documentclass[twocolumn]{article}
\usepackage{epsfig}
\usepackage{graphicx}
\usepackage{color}
\usepackage[acmtitlespace,nocopyright]{acmneww}
\usepackage{times}
\usepackage{amsmath}
\usepackage{amssymb}
\usepackage{xspace}
\usepackage{subfigure}

\newcommand {\mm}[1] {\ifmmode{#1}\else{\mbox{\(#1\)}}\fi}

\newcommand{\ignore}[1]{}

\newcommand{\proof}{\noindent{\sc Proof.~}}
\newcommand{\eop}{\hfill\usebox{\smallProofsym}\bigskip}  %
\newsavebox{\smallProofsym}                            % smallproofsymbol
\savebox{\smallProofsym}                               %
{%                                                     %
\begin{picture}(6,6)                                   %
\put(0,0){\framebox(6,6){}}                            %
\put(0,2){\framebox(4,4){}}                            %
\end{picture}                                          %
}                                                      %
\makeatletter
\long\def\@makecaption#1#2{%
  \vskip\abovecaptionskip
  \sbox\@tempboxa{\small #1: #2}%
  \ifdim \wd\@tempboxa >\hsize
    \small #1: #2\par
  \else
    \global \@minipagefalse
    \hb@xt@\hsize{\hfil\box\@tempboxa\hfil}%
  \fi
  \vskip\belowcaptionskip}
\makeatother

\newcommand{\uuu}           {\mm{\tt u}}
\newcommand{\vvv}           {\mm{\tt v}}
\newcommand{\UU}            {\mm{U}}
\newcommand{\Att}           {\mm{\bar{\mathcal A}}}
\newcommand{\Btt}           {\mm{\bar{\mathcal B}}}
\newcommand{\Dtt}           {\mm{\bar{\mathcal D}}}
\newcommand{\Utt}           {\mm{\tt U}}

\newcommand{\bbb}           {\mm{\bf B}}
\newcommand{\ccc}           {\mm{\bf C}}
\newcommand{\rrr}           {\mm{\bf R}}
\newcommand{\AAA}           {\mm{\mathcal A}}
\newcommand{\BBB}           {\mm{\mathcal B}}
\newcommand{\DDD}           {\mm{\mathcal D}}
\newcommand{\RRR}           {\mm{\mathcal R}}
\newcommand{\VVV}           {\mm{\mathcal V}}
\newcommand{\voronoi}[1]    {\mm{\rm vor}{({#1})}}
\newcommand{\restricted}[1] {\mm{\rm res}{({#1})}}

\newcommand{\ssx}           {\mm{\sigma}}

\newcommand{\Bgroup}        {\mm{\sf B}}
\newcommand{\Cgroup}        {\mm{\sf C}}
\newcommand{\Fgroup}        {\mm{\sf F}}

\newcommand{\Hgroup}        {\mm{\sf H}}
\newcommand{\Rgroup}        {\mm{\sf R}}
\newcommand{\Vgroup}        {\mm{\sf V}}
\newcommand{\Zgroup}        {\mm{\sf Z}}

\newcommand{\Betti}         {\mm{\beta}}

\newcommand{\Ddgm}          {\mm{\rm Dgm}}
\newcommand{\Odgm}          {\mm{\rm Ord}}
\newcommand{\Hdgm}          {\mm{\rm Hor}}
\newcommand{\Vdgm}          {\mm{\rm Vcl}}
\newcommand{\Rdgm}          {\mm{\rm Rel}}
\newcommand{\norm}[1]       {\mm{\|{#1}\|}}
\newcommand{\kernel}[1]     {\mm{\sf ker\,}{#1}}
\newcommand{\image}[1]      {\mm{\sf im\,}{#1}}

\newcommand{\rank}[1]       {\mm{\sf rank\,}{#1}}
\newcommand{\pers}[1]       {\mm{\rm pers{({#1})}}}

\newcommand{\Rspace}        {\mm{{\mathbb R}}}

\newcommand{\Edist}[2]      {\mm{\|{#1}-{#2}\|}}

\newcommand{\capsp}         {{\; \cap \;}}

\newtheorem{result}{}

\title{The Medusa of Spatial Sorting: ~\\ Topological Construction
       \thanks{This research is partially supported
               by NSF under grant DBI-0820624,
               by ESF under the Research Network Programme,
               and by the Russian Government under mega project 11.G34.31.0053.}
       }

\author{Herbert Edelsbrunner\thanks{IST Austria (Institute of Science and
            Technology Austria), Kloster\-neu\-burg, Austria;
            Departments of Computer Science and of Mathematics,
            Duke University, Durham, North Carolina;
            Delaunay Lab of Discrete and Computational Geometry,
            Yaroslavl' State University, Russia;
            Geomagic, Research Triangle Park, North Carolina.},
        Carl-Philipp Heisenberg${}^\ddagger$,
        Michael Kerber${}^\ddagger$ and
        Gabriel Krens\thanks{IST Austria (Institute of Science and
            Technology Austria), Kloster\-neu\-burg, Austria.}}

\begin{document}
\maketitle

\begin{abstract}
  We consider the simultaneous movement of finitely many colored points
  in space, calling it a \emph{spatial sorting process}.
  The name suggests a purpose that drives the collection to
  a configuration of increased or decreased order.
  Mapping such a process to a subset of space-time,
  we use persistent homology measurements of the time function
  to characterize the process topologically.
\end{abstract}

\vspace{0.1in}

%\clearpage
%%%%%%%%%%%%%%%%%%%%%%%%%%%%%%%%%%%%%%%%%%%%%%%%%%%%%%%%%%%%%%%%%%%%%%%%%%
%%%%%%%%%%%%%%%%%%%%%%%%%%%%%%%%%%%%%%%%%%%%%%%%%%%%%%%%%%%%%%%%%%%%%%%%%%
\section{Introduction}
\label{sec1}
%%%%%%%%%%%%%%%%%%%%%%%%%%%%%%%%%%%%%%%%%%%%%%%%%%%%%%%%%%%%%%%%%%%%%%%%%%
%%%%%%%%%%%%%%%%%%%%%%%%%%%%%%%%%%%%%%%%%%%%%%%%%%%%%%%%%%%%%%%%%%%%%%%%%%

Motivated by the observation of cell communities over time,
we propose a topological expression of the process
that facilitates the identification and quantification of
its features.
In this manuscript, we focus on the mathematical framework.

\paragraph{Motivation.}
The specific motivation for the work described in this paper
is the experimental data on cell segregation in the developing
zebrafish imaged by Heisenberg and Krens \cite{HeKr11}.
Making cells of two populations observable through fluorescent markers,
they follow them through time,
assigning each population (cell type) its own unique color.
In this particular case, the two populations start spatially mixed
and end in spatially segregated configurations.
The segregation is captured by a series of $3$-dimensional images,
which we turn into a shape in space-time.

Spatial segregation is a special case of the
broader class of \emph{spatial sorting processes},
in which we are given one or more distinguishable populations of
particles (points in space),
and we are interested in characterizing their spatial rearrangement in time.
We aim at characterizing the spatial sorting process through
detailed measurements of its features.
The quantification may be used to establish a classification of
spatial sorting processes or, on a finer scale,
to differentiate between realizations of the same process.
A common biological application is the establishment of phenotypes
that can help in the classification of genetic influences. 
Once we have a description of the process beyond initial and final states,
we may ask more subtle questions, such as whether an observed inverse
process has a symmetric characterization.

\paragraph{Results.}
Our contributions are primarily mathematical,
with the goal of using the insights toward the quantitative
analysis of experimental time-series data:
\begin{enumerate}
  \item[(i)]  we \emph{model} a spatial sorting process as a shape in
    space-time with descriptive topological properties;
  \item[(ii)]  we \emph{measure} this shape using the
    persistent homology of the time function;
  \item[(iii)]  we \emph{provide} a classification of the measurements,
    interpreting them as aspects of the process.
\end{enumerate}
Note that measuring the process in space-time is different
from taking the trajectory of measurements of the sequence of time-slices.
Indeed, we will distinguish between
\emph{temporary} space-time features, that can be observed in slices,
from \emph{fleeting} features that cannot be so observed.
The latter require memory and temporal reasoning
and are therefore less readily accessible to an observer who lives in time.

The main idea of our approach is to turn the time-series of
geometric data into a $4$-dimensional topological space
whose connectivity is descriptive of the spatial sorting process.
The construction takes three steps to produce the measurements:
\begin{description}
  \item[{\sc Step 1:}]  construct the Voronoi tessellation to turn
    the data in a time-slice into a subset of $\Rspace^3$;
  \item[{\sc Step 2:}]  maintain the construction through time,
    effectively sweeping out a subset of $\Rspace^3 \times [0,1]$;
  \item[{\sc Step 3:}]  measure the constructed subset of space-time
    using the persistent homology of its time function.
\end{description}
We explain the first two steps in Section \ref{sec2} and the
third step in the Section \ref{sec3}.
Details of the corresponding algorithms and their implementations
can be found in \cite{KeEd12}.
Section \ref{sec4} discusses the information provided by persistent homology.
Section \ref{sec5} illustrates the ideas by running the corresponding
algorithms on simulated data.
Section \ref{sec6} concludes the paper.

%\clearpage
%%%%%%%%%%%%%%%%%%%%%%%%%%%%%%%%%%%%%%%%%%%%%%%%%%%%%%%%%%%%%%%%%%%%%%%%%%
%%%%%%%%%%%%%%%%%%%%%%%%%%%%%%%%%%%%%%%%%%%%%%%%%%%%%%%%%%%%%%%%%%%%%%%%%%
\section{Geometry}
\label{sec2}
%%%%%%%%%%%%%%%%%%%%%%%%%%%%%%%%%%%%%%%%%%%%%%%%%%%%%%%%%%%%%%%%%%%%%%%%%%
%%%%%%%%%%%%%%%%%%%%%%%%%%%%%%%%%%%%%%%%%%%%%%%%%%%%%%%%%%%%%%%%%%%%%%%%%%

In this section, we introduce the sets and complexes
we use to model spatial sorting processes.
In Euclidean space, we call them \emph{tessellations},
\emph{triangulations}, and \emph{complexes},
and in space-time, we call them \emph{medusas}.
They all consist of \emph{cells} of various dimensions,
but because we also talk about cells in the biological sense,
we will use special names, such as \emph{Voronoi cells} and
\emph{simplices} in Euclidean space, and \emph{stacks}
and \emph{prisms} in space-time, whenever possible.

%%%%%%%%%%%%%%%%%%%%%%%%%%%%%%%%%%%%%%%%%%%%%%%%%%%%%%%%%%%%%%%%%%%%%
\subsection{A Moment in Time}
\label{sec21}
%%%%%%%%%%%%%%%%%%%%%%%%%%%%%%%%%%%%%%%%%%%%%%%%%%%%%%%%%%%%%%%%%%%%%

The input data is a finite set of colored points in $\Rspace^3$.
Writing the color as subscript, we let $\UU$ be the union
of $\UU_1$ to $\UU_k$, assuming $\UU_i \capsp \UU_j = \emptyset$ for all $i \neq j$.

\paragraph{Voronoi tessellation and Delaunay triangulation.}
The \emph{Voronoi cell} of $u \in \UU$
consists of all $x \in \Rspace^3$ for which $u$ minimizes
the Euclidean distance among all points in $\UU$:
\begin{eqnarray}
  \voronoi{u}  &=&  \{ x \in \Rspace^3 \mid
                    \Edist{x}{u} \leq \Edist{x}{v}, v \in \UU \} .
\end{eqnarray}
The set of Voronoi cells, $V = V(\UU) = \{ \voronoi{u} \mid u \in \UU \}$,
is the \emph{Voronoi tessellation} of $\UU$.
For each $1 \leq \ell \leq k$, we write
$V_\ell = \{ \voronoi{u} \mid u \in \UU_\ell \}$ for the subset
of Voronoi cells of that color.
Note that $V$ is the disjoint union of $V_1, V_2, \ldots, V_k$.
\begin{figure}[hbt]
 \vspace{0.0in}
 \centering
 \resizebox{!}{1.8in}{\input{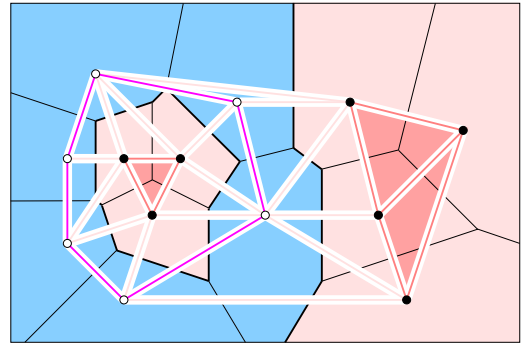}}
 \caption{The full subcomplexes defined by the two colors of a
   bi-chromatic Delaunay triangulation in the Euclidean plane.}
 \label{fig:Delaunay}
\end{figure}
It is often convenient to work with the dual instead of directly
with the Voronoi tessellation.
Abstractly, this is the \emph{nerve} of the set of Voronoi cells,
that is: the system of subsets whose Voronoi cells have a non-empty
common intersection.
Assuming general position of the points in $\UU$,
the number of Voronoi cells that can have a non-empty common
intersection is at most $4$.
We represent each set in the nerve by the convex hull
of the points that generate its Voronoi cells,
which can be a vertex, and edge, a triangle, or a tetrahedron.
Together, these convex hulls form a simplicial complex,
known as the \emph{Delaunay triangulation} of $\UU$,
and denoted as $D = D(\UU)$.
For each $1 \leq \ell \leq k$, we write $D_\ell$ for the
full subcomplex of $D$ that consists of all vertices of color $\ell$
and all edges, triangles, and tetrahedra that connect them.
As illustrated in Figure \ref{fig:Delaunay} for $\Rspace^2$,
the full subcomplexes of different colors are disjoint,
with multi-colored edges, triangles, and tetrahedra between them.

\begin{figure}[hbt]
 \vspace{0.0in}
 \centering
 \resizebox{!}{1.8in}{\input{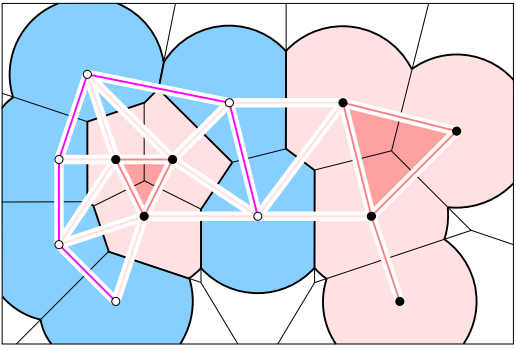}}
 \caption{The full subcomplexes defined by the two colors of a
   bi-chromatic alpha complex in the Euclidean plane.}
 \label{fig:alpha}
\end{figure}

\paragraph{Restricted Voronoi tessellation and alpha complex.}
In a loose configuration, a biological cell would generally not occupy
the entire space alloted to it by the Voronoi cell of its nucleus.
To better approximate the space used by the cell, we therefore
choose a fixed positive radius, $\alpha_0$,
and consider the \emph{restriction} of the Voronoi cell
to the ball centered at the generating point:
\begin{eqnarray}
  \restricted{u}  &=&  \{ x \in \voronoi{u} \mid \Edist{x}{u} \leq \alpha_0 \} .
\end{eqnarray}
Similar to before, we write $R = R(\UU)$ for the set of
restricted Voronoi cells, and $R_\ell \subseteq R$ for the subset
of cells generated by points of color $\ell$.
Clearly, $R$ is the disjoint union of $R_1, R_2, \ldots, R_k$.
Note that $\restricted{u} \subseteq \voronoi{u}$ for each point $u \in \UU$.
It follows that the nerve of $R$ is isomorphic to a subsystem of the nerve of $V$.
Accordingly, we define the dual \emph{alpha complex} to consist
of the simplices in $D$ whose corresponding restricted Voronoi cells
have a non-empty common intersection,
denoting it as $A = A(\UU)$.
For each $1 \leq \ell \leq k$, we again have the full subcomplex
$A_\ell \subseteq A$ that contains all vertices of color $\ell$ and all
simplices connecting them.
Figure \ref{fig:alpha} illustrates the definitions by restricting
the Voronoi cells in Figure \ref{fig:Delaunay} to their disks.
Note that $A_\ell$ is also a subcomplex of $D$, but not necessarily
a full subcomplex.
Indeed, we have $A_\ell = D_\ell \capsp A$, so $A_\ell$ is a full
subcomplex of $D$ iff $A_\ell = D_\ell$.

\paragraph{Homotopy equivalence.}
An important structural relationship between unions of
Voronoi cells and their dual complexes follows from
the Nerve Theorem; see e.g.\ \cite[page 59]{EdHa10}.
To state the relationship, we ignore the difference between a set
of Voronoi cells and their union, as well as the difference
between a simplicial complex and its underlying space.
We write $X \simeq Y$ if the sets $X$ and $Y$ have the same homotopy type;
see \cite[page 108]{Mun84} for a definition.
\begin{result}[Lemma A]
  We have $R_\ell \simeq A_\ell$, for each color $1 \leq \ell \leq k$.
\end{result}
\proof
 All cells in $R_\ell$ are closed and convex.
 It therefore suffices to use the corresponding special version 
 of the Nerve Theorem.
 It says that the union of the sets and the nerve
 of the collection have the same homotopy type.
 The claimed relationships follow because $A_\ell$ is
 a geometric realization of the nerves of $R_\ell$.
\eop

We get $V_\ell \simeq D_\ell$ as a special case.
Lemma A implies that $R_\ell$ and $A_\ell$ have isomorphic homology groups.
Instead of dealing with the computationally more difficult
union of Voronoi cells, we can therefore compute the homology for the dual complex,
which is a purely combinatorial object.

%%%%%%%%%%%%%%%%%%%%%%%%%%%%%%%%%%%%%%%%%%%%%%%%%%%%%%%%%%%%%%%%%%%%%
\subsection{Trajectories}
\label{sec22}
%%%%%%%%%%%%%%%%%%%%%%%%%%%%%%%%%%%%%%%%%%%%%%%%%%%%%%%%%%%%%%%%%%%%%

Assuming a continuous trajectory for each data point, we form subsets
of space-time by taking unions of Voronoi cells,
both in space and in time.
We begin with some definitions.
A \emph{trajectory} is a continuous mapping
$\uuu: [a,b] \to \Rspace^3$, with $0 \leq a < b \leq 1$.
We let $\Utt$ be a finite set of trajectories,
assuming no two intersect in space-time;
that is: $\vvv (t) \neq \uuu (t)$ for all $\vvv \neq \uuu$ in $\Utt$
and all $t \in [0,1]$ for which both trajectories are defined.
Writing $[k]$ for the set $\{1, 2, \ldots, k\}$,
we let $\chi: \Utt \to [k]$ be a coloring.
At each time $t \in [0,1]$, we have a finite set of points,
$\Utt (t) = \{ \uuu (t) \mid \uuu \in \Utt \}$,
of course taking only the trajectories with $a \leq t \leq b$.
The finite set of points is also colored,
with coloring induced by $\chi$.

\begin{figure}[hbt]
 \vspace{0.0in}
 \centering
 %% \resizebox{!}{1.8in}{\input{Figs/stacks.pstex_t}}
 \centerline{\epsfig{figure=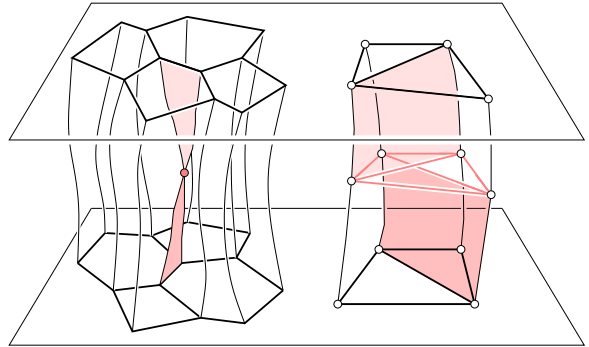,height=1.8in}}
 \caption{Left: four stack of $2$-dimensional Voronoi cells.
   Right: the four corresponding prisms in the Delaunay medusa
   connected by a $3$-simplex representing the flip between
   the two ways to triangulate four points.}
 \label{fig:stacks}
\end{figure}

\paragraph{Voronoi and Delaunay medusas.}
For each point $\uuu (t) \in \Utt (t)$,
we write $\voronoi{\uuu (t)}$ for its Voronoi cell in $\Rspace^3 \times t$.
The Voronoi tessellation at time $t$ is denoted as $V(t) = V( \Utt (t))$,
and the subset of Voronoi cells of color $\ell$ is denoted as
$V_\ell (t) \subseteq V(t)$.
Collecting Voronoi cells in time,
we get a $1$-parameter family
of cells generated by a trajectory:
\begin{eqnarray}
  \voronoi{\uuu}  &=&  \bigcup_{t \in [a,b]} \voronoi{\uuu (t)} .
  \label{eqn:Voronoi}
\end{eqnarray}
Noting that the Voronoi cells on the right hand side of \eqref{eqn:Voronoi}
lie in distinct parallel copies of $\Rspace^3$,
we call $\voronoi{\uuu}$ a \emph{stack}.
While the Voronoi cell in each time-slice is
a $3$-dimensional convex polyhedron,
the stack itself is neither necessarily convex nor necessarily polyhedral;
see Figure \ref{fig:stacks} on the left.
Switching fonts, we write $\VVV = \VVV (\Utt)$ for the set of stacks,
each defined by a trajectory in $\Utt$.
For each $1 \leq \ell \leq k$, we write $\VVV_\ell = \VVV_\ell (\Utt)$
for the subset of stacks generated by trajectories of color $\ell$.
We call $\VVV$ the \emph{multi-chromatic}
and each $\VVV_\ell$ a \emph{mono-chromatic Voronoi medusa}.

Similar to stacks of Voronoi cells, we also consider stacks
of Delaunay simplices, which we call \emph{prisms}.
There is an important difference caused by the occasional
sudden change of the Delaunay triangulation.
Such a change is a \emph{flip}, which either replaces two tetrahedra
by three, or three tetrahedra by two; see e.g.\ \cite[page 102]{Ede01}.
In $4$-dimensional space-time, a flip appears as a (degenerate) $4$-simplex
that connects to the preceding Delaunay triangulations along two tetrahedra
and to the succeeding Delaunay triangulations along three tetrahedra,
or the other way round.
Reducing the insertion or deletion of a point to a sequence of flips,
as described in \cite[Section 5.3]{Ede01}, we see that prisms and
$4$-simplices suffice to fully describe the history
of the Delaunay triangulation.
We write $\DDD = \DDD (\Utt)$ for the complex in $\Rspace^3 \times [0,1]$
and $\DDD_\ell = \DDD_\ell (\Utt)$ for the full subcomplex
of color $\ell$, calling $\DDD$ the \emph{multi-chromatic}
and $\DDD_\ell$ a \emph{mono-chromatic Delaunay medusa}.
Figure \ref{fig:stacks} illustrates the definitions
in $\Rspace^2 \times [0,1]$.

\paragraph{Restricted Voronoi and alpha medusas.}
Following the distinction between unrestricted
and restricted Voronoi cells, we extend the latter to
space-time in the obvious way.
We thus introduce a \emph{stack} of restricted Voronoi cells,
\begin{eqnarray}
  \restricted{\uuu}  &=&  \bigcup_{t \in [a,b]} 
                          \restricted{\uuu (t)} , 
\end{eqnarray}
the set of such stacks, $\RRR = \RRR (\Utt)$,
and the complex of prisms and $4$-simplices swept out by the simplices
in the alpha complex, $\AAA = \AAA (\Utt)$.
Furthermore, we introduce the colored subsets, $\RRR_\ell \subseteq \RRR$,
and the colored subcomplexes, $\AAA_\ell \subseteq \AAA$,
with $\ell \in [k]$.
We call $\RRR$ the \emph{multi-chromatic}
and each $\RRR_\ell$ a \emph{mono-chromatic restricted Voronoi medusa}.
Similarly, we call $\AAA$ the \emph{multi-chromatic}
and each $\AAA_\ell$ a \emph{mono-chromatic alpha medusa}.

\paragraph{Homotopy equivalence.}
The structural relationship between unions of 
Voronoi cells and their dual complexes extends from
$\Rspace^3$ to $\Rspace^3 \times [0,1]$.
As before, we simplify the notation by ignoring the difference
between a collection of cells and its union.
\begin{result}[Lemma B]
  We have $\RRR_\ell \simeq \AAA_\ell$, for each color $\ell \in [k]$.
\end{result}
\proof
 As in the proof of Lemma A,
 we appeal to the Nerve Theorem to give the claimed relationships.
 Since the cells are no longer convex, we use the more general
 version that applies to finite collections of closed, contractible
 sets whose intersections (of any order) are again contractible.

\begin{figure}[hbt]
  \vspace{0.0in}
  \centering
  \centerline{\epsfig{figure=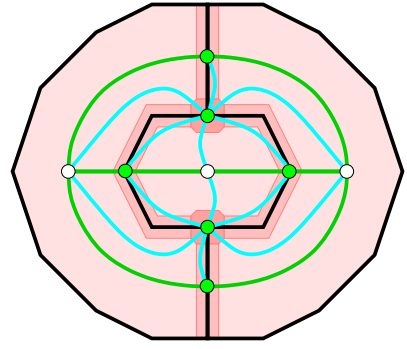,height=1.8in}}
  \caption{Three $2$-dimensional stacks (lightly shaded)
    giving rise to a decomposition into nine bodies (shaded).
    The nerve of the bodies forms the barycentric subdivision of
    two triangles.}
  \label{fig:nerve}
 \end{figure}

 While each stack in $\RRR_\ell$ of any dimension is swept out
 by a convex set and is therefore contractible,
 two stacks can intersect in two or more (lower-dimensional) stacks,
 which prevents the direct application of the Nerve Theorem.
 We finesse the difficulty by shrinking each stack at its boundary
 and thickening it into a $4$-dimensional body;
 see Figure \ref{fig:nerve}.
 There is more than one way to do this such that the bodies are
 closed and contractible, their union is the same as the union of
 the original stacks, and the common intersection of $i+1$ bodies
 is either empty or a contractible $(4-i)$-dimensional set.
 For example, we may use the mixed complex defined for Voronoi
 cells in \cite{Ede99} to sweep out the bodies.
 Letting $\BBB_\ell$ be the resulting set of bodies,
 we consider the nerve, which we denote by $\Btt_\ell$.
 As illustrated in Figure \ref{fig:nerve}, this nerve is the barycentric
 subdivision of a complex of simplices whose vertices correspond to
 the original $4$-dimensional stacks.
 Note that this complex is \emph{not} necessarily a simplicial complex,
 since two simplices may share more than just one common face.
 Denoting this complex of simplices by $\Att_\ell$, we have
 \begin{eqnarray}
   \RRR_\ell ~~\simeq~~ \BBB_\ell ~~\simeq~~ \Btt_\ell ~~\simeq~~ \Att_\ell,
 \end{eqnarray}
 in which the only non-trivial, middle relation
 is furnished by the Nerve Theorem.
 To complete the argument, we still need to show that $\Att_\ell$
 and $\AAA_\ell$ have the same homotopy type.
 To do this, we contract each prism in $\AAA_\ell$ along the
 time direction.
 Indeed, each prism is a simplex times a time interval, and
 the contraction glues the bottom to the top face,
 effectively turning the prism into a simplex.
 We do this for all prisms simultaneously in order to avoid temporary
 inconsistencies between the prisms and their side faces,
 which are lower-dimensional prisms.
 Eventually, when all prisms have be contracted,
 we have turned the medusa into $\Att_\ell$.
 Since the contraction preserves the homotopy type,
 this implies the claimed relation.
\eop

We get $\VVV_\ell \simeq \DDD_\ell$ as a special case.
Lemma B also implies that we get isomorphic homology groups and can
therefore do the computation on the dual medusa, or indeed
the complex of simplices, $\Att_\ell$, which is a purely combinatorial object.
We refer to it as the \emph{data structure} for the mono-chromatic alpha medusa;
beyond being instrumental in the proof of Lemma B,
it is the main computational tool we use to implement
that algorithms that are only sketched in this paper;
see \cite{KeEd12}.

%%%%%%%%%%%%%%%%%%%%%%%%%%%%%%%%%%%%%%%%%%%%%%%%%%%%%%%%%%%%%%%%
\subsection{Filtered Sequences}
\label{sec23}
%%%%%%%%%%%%%%%%%%%%%%%%%%%%%%%%%%%%%%%%%%%%%%%%%%%%%%%%%%%%%%%%

We use the restricted Voronoi medusa to explain
filtered sequences of spaces,
and free up to subscript by focusing the discussion on the
multi-chromatic version.

\paragraph{Sub- and superlevel sets.}
The appropriately restricted \emph{time function},
$f|_\RRR: \RRR \to [0,1]$,
maps every point $x \in \RRR$ to its time coordinate, $f|_\RRR (x)$.
Given a moment in time, $t \in [0,1]$, the corresponding
\emph{sublevel set} is $\RRR_t = f|_\RRR^{-1} [0,t]$,
and the corresponding \emph{superlevel set}
is $\RRR^t = f|_\RRR^{-1} [t,1]$.
We will be interested in the filtered sequences:
\begin{eqnarray}
  \emptyset \subseteq \RRR_0 \subseteq \ldots
            \subseteq \RRR_t \subseteq \ldots \subseteq \RRR_1 = \RRR ,               
                                                 \label{eqn:sublevel}  \\
  \emptyset \subseteq \RRR^1 \subseteq \ldots
            \subseteq \RRR^t \subseteq \ldots \subseteq \RRR^0 = \RRR .
                                                 \label{eqn:superlevel}
\end{eqnarray}
They capture the evolution of the restricted Voronoi medusa in a way that
allows us to compare different states without the burden of
computing a correspondence.
Similarly, we consider the time function restricted to the
alpha medusa, define sub- and superlevel sets,
$\AAA_t$ and $\AAA^t$, and form filtered sequences as before.
Importantly, the transformation from the stacks to the prisms
does not change the homotopy type.
\begin{result}[Lemma C]
  We have $\RRR_t \simeq \AAA_t$ as well as
  $\RRR^t \simeq \AAA^t$, for all $t \in [0,1]$.
\end{result}
We omit the proof which is similar to that of Lemma B.
It is important to realize that Lemma C extends to all mono-chromatic
medusas.

\paragraph{Complexes of simplices.}
Following the proof of Lemma B, we find it convenient to replace
the sub- and superlevel sets of the alpha medusa by complexes of simplices.
Specifically, we write $\Att_t$ for the complex
obtained by contracting all prisms in $\AAA_t$ in time direction.
Symmetrically, we write $\Att^t$ for the complex
obtained by contracting all prisms in $\AAA^t$.
There is an alternative description.
Let $\varphi$ be a prism in $\AAA$, and write $f_{\min} (\varphi)$
and $f_{\max} (\varphi)$ for the minimum and maximum time values
of the points in $\varphi$.
We assign these two values as $\ssx_{\min} \leq \ssx_{\max}$
to the simplex $\ssx$ in $\Att$ obtained by contracting $\varphi$.
Then $\Att_t$ is the subcomplex of simplices with $\ssx_{\min} \leq t$,
and $\Att^t$ is the subcomplex of simplices with $\ssx_{\max} \geq t$.
Note that $\Att_t$ and $\Att^t$ overlap in the simplices
that correspond to the prisms with
$f_{\min} (\varphi) \leq t \leq f_{\max} (\varphi)$.
Similar to the sub- and superlevel sets, the complexes
of simplices form filtered sequences:
\begin{eqnarray}
  \emptyset \subseteq \Att_0 \subseteq
     \ldots \subseteq \Att_t \subseteq
     \ldots \subseteq \Att_1 = \Att ,               
                                               \label{eqn:subcplx}  \\
  \emptyset \subseteq \Att^1 \subseteq
     \ldots \subseteq \Att^t \subseteq
     \ldots \subseteq \Att^0 = \Att ,
                                               \label{eqn:supercplx}
\end{eqnarray}
Again, the transformation does not affect the homotopy type.
Indeed, the complexes forming the filtered sequences in
\eqref{eqn:subcplx} and \eqref{eqn:supercplx} are reminiscent of
the lower- and upper-star filtrations we find in \cite{EdHa10}.

%\clearpage
%%%%%%%%%%%%%%%%%%%%%%%%%%%%%%%%%%%%%%%%%%%%%%%%%%%%%%%%%%%%%%%%%%%%%%%%%%
%%%%%%%%%%%%%%%%%%%%%%%%%%%%%%%%%%%%%%%%%%%%%%%%%%%%%%%%%%%%%%%%%%%%%%%%%%
\section{Algebra}
\label{sec3}
%%%%%%%%%%%%%%%%%%%%%%%%%%%%%%%%%%%%%%%%%%%%%%%%%%%%%%%%%%%%%%%%%%%%%%%%%%
%%%%%%%%%%%%%%%%%%%%%%%%%%%%%%%%%%%%%%%%%%%%%%%%%%%%%%%%%%%%%%%%%%%%%%%%%%

In this section, we turn the spaces of Section \ref{sec2} into
algebraic information.
The foundation of the transformation is the classical notion of homology,
which we review.
The information is summarized in persistence diagrams,
which we introduce for modules obtained from sub- and superlevel sets
of the time function.

%%%%%%%%%%%%%%%%%%%%%%%%%%%%%%%%%%%%%%%%%%%%%%%%%%%%%%%%%%%%%%%%
\subsection{Measuring Connectivity}
\label{sec31}
%%%%%%%%%%%%%%%%%%%%%%%%%%%%%%%%%%%%%%%%%%%%%%%%%%%%%%%%%%%%%%%%

Homology groups detect and count holes in a single space.
We begin with a brief introduction of this classical subject;
see \cite{Mun84} for more information.

\paragraph{Homology groups.}
Consider a simplicial complex, perhaps an alpha complex, $A$,
which consists of simplices of dimension $0 \leq p \leq 3$.
Fixing a field of coefficients, $\Fgroup$, we call a formal
sum of the form $\gamma = \sum c_i \ssx_i$ a \emph{$p$-chain},
in which the $c_i$ are elements of $\Fgroup$ and the $\ssx_i$
are $p$-simplices in $A$, each with a fixed orientation.
The \emph{boundary} of the $p$-chain is the similarly weighted
sum of boundaries of the simplices:
$\partial_p \gamma = \sum c_i \partial_p \ssx_i$,
in which $\partial_p \ssx_i$ is the sum of the $(p-1)$-simplices
that are its faces.
We call $\gamma$ a \emph{$p$-cycle} if $\partial_p \gamma = 0$,
and we call $\gamma$ a \emph{$p$-boundary} if there is
a $(p+1)$-chain $\delta$ with $\gamma = \partial_{p+1} \delta$.
The chains thus form \emph{chain groups} connected by
boundary homomorphisms, $\partial_p : \Cgroup_p \to \Cgroup_{p-1}$.
Similarly, the cycles form \emph{cycle groups}, the kernels of
the boundary homomorphisms, $\Zgroup_p = \kernel{\partial_p}$,
and the boundaries form \emph{boundary groups}, 
the images of the boundary homomorphisms,
$\Bgroup_p = \image{\partial_{p+1}}$.
Since the boundary of a boundary is necessarily zero,
we have $\Bgroup_p \subseteq \Zgroup_p$ and we can take the quotient,
$\Hgroup_p = \Zgroup_p / \Bgroup_p$,
which is the \emph{$p$-th homology group}.
Homology groups can be defined quite generally,
for example, as explained above for triangulations of topological spaces.
Since we choose the coefficients from a field, all groups
we mentioned are vector spaces, which are characterized by their ranks.
For the $p$-th homology group, the rank is called
the \emph{$p$-th Betti number}, denoted as $\Betti_p = \rank{\Hgroup_p}$.

For a space of dimension $k$,
the only possibly non-zero Betti numbers are $\Betti_0$ to $\Betti_k$.
In our case, $A$ has dimension at most $k = 3$,
and we have $\Betti_3 = 0$ because
every $3$-chain in $\Rspace^3$ has non-zero boundary.
The remaining three possibly non-zero Betti numbers
have intuitive interpretations:
$\Betti_0$ counts components,
$\Betti_1$ counts loops, and
$\Betti_2$ counts completely surrounding walls.
We get additional intuition by observing that the connectivity
of the complement space, $\Rspace^3 - A$, depends on the
connectivity of $A$,
a relation formalized by Alexander Duality \cite[p.\ 424]{Mun84}.
We refer to the elements of the homology group of the complement
space as the \emph{holes} of $A$.
Distinguishing between the different dimensions,
$\Betti_0 - 1$ counts \emph{gaps} between the components,
$\Betti_1$ counts \emph{tunnels} passing through the loops, and
$\Betti_2$ counts \emph{voids} surrounded by walls.
We will compute Betti numbers for medusas in $\Rspace^3 \times [0,1]$.
While more complicated than in $3$-dimensional space,
we can still interpret the Betti numbers in terms of \emph{evolutions}
of gaps, tunnels, and voids, as we will explain in Section \ref{sec4}.

In addition to single spaces, we take the homology of pairs
in which the second space
is a subset of the first, such as $(\RRR, \RRR^t)$, for example.
The \emph{$p$-th relative homology group}, denoted as
$\Hgroup_p (\RRR, \RRR^t)$,
is defined the same way as $\Hgroup_p (\RRR)$ except
that differences in $\RRR^t$ are ignored.
In other words, two chains are the same if they agree on
all simplices in $\RRR - \RRR^t$, etc.
To get an intuition for these groups, we may
add the cone over $\RRR^t$.
Then the rank of $\Hgroup_p (\RRR, \RRR^t)$ is the same
as the rank of the $p$-th absolute homology group
for the modified complex, except for $p=0$, where it is one less.

\paragraph{Ordinary and extended persistence modules.}
In the language of category theory, homology is a functor that
maps a space, or a pair of spaces, to a sequence of groups,
one for each dimension.
To simplify the notation, we write $\Hgroup = \bigoplus_p \Hgroup_p$
for the direct sum of the individual groups.
This frees up the subscript, which we use to index the homology
groups of a filtered sequence.
Letting $0 = t_1 < t_2 < \ldots < t_m = 1$
be the homological critical values of $f_\RRR$,
we note that the homology group is constant
for all $t_i \leq t < t_{i+1}$.
Hence, it suffices to list the finitely many groups
$\Rgroup_i = \Hgroup (\RRR_{t_i})$ to describe the entire
$1$-parameter family,
giving a finite sequence of homology groups
connected by homomorphisms induced by inclusion:
\begin{eqnarray}
  0 ~=~ \Rgroup_0 \to \Rgroup_1 \to \ldots
                  \to \Rgroup_m ~=~ \Hgroup (\RRR) .
  \label{eqn:ordinary}
\end{eqnarray}
We call \eqref{eqn:ordinary} a \emph{persistence module}, or more specifically
the \emph{ordinary persistence module} of $f$.
As described in \cite[Section VII.1]{EdHa10},
we can use the persistence module to define \emph{births} and \emph{deaths}
of homology classes and arrange them in pairs.
Specifically, a class is \emph{born} at $\Rgroup_i$ if it does not
belong to the image of $\Rgroup_{i-1}$,
and it \emph{dies entering} $\Rgroup_j$ if $j$ is the smallest index
for which the image of the class lies in the image of $\Rgroup_{i-1}$.
The birth-death pair describes a coset of homology classes and specifies
their \emph{persistence} as the absolute difference between the function
values at the two events, in this case $t_j - t_i$.

The ordinary module \eqref{eqn:ordinary} allows for homology classes
that are born but never die.
These are the classes that describe the entire medusa, $\RRR$,
which are of particular interest to us.
To get a finite measurement of duration,
we extend the module using the relative homology groups of the form
$\Hgroup (\RRR, \RRR^t)$.
Writing $\Rgroup_{m+i} = \Hgroup (\RRR, \RRR^{t_{m-i+1}})$, we get
\begin{eqnarray}
  0 ~=~ \Rgroup_0 \to \ldots \to
        \Rgroup_m \to \ldots \to \Rgroup_{2m} ~=~ 0 ,
  \label{eqn:extended}
\end{eqnarray}
which we call the \emph{extended persistence module} of $f|_\RRR$.
It begins and ends with the trivial group,
which implies that everything that is born also dies.
We thus get a more complete set of measurements of the alpha
medusa from \eqref{eqn:extended} than from \eqref{eqn:ordinary}.

\paragraph{Isomorphic relative homology groups.}
Lemma C implies that the (absolute) homology groups of the
sublevel sets of the restricted Voronoi medusa are
isomorphic to those of the alpha medusa.
We now show that the same is true for the relative homology groups.
\begin{result}[Lemma D]
  The relative homology groups $\Hgroup (\RRR, \RRR^t)$ and
  $\Hgroup (\AAA, \AAA^t)$
  are isomorphic, for all $t \in [0,1]$.
\end{result}
\proof
 Recall the exact sequences of the two pairs, which we write
 from left to right and in parallel:
 $$
  \begin{array}{ccccc}
    \hspace{-0.0cm}    \Hgroup_p     (\AAA^t)       \hspace{-0.05cm} \to
      & \hspace{-0.35cm}\Hgroup_p     (\AAA)        \hspace{-0.05cm} \to
      & \hspace{-0.35cm}\Hgroup_p     (\AAA, \AAA^t)\hspace{-0.05cm} \to
      & \hspace{-0.35cm}\Hgroup_{p-1} (\AAA^t)      \hspace{-0.05cm} \to
      & \hspace{-0.35cm}\Hgroup_{p-1} (\AAA)                             \\
    \hspace{-0.4cm} \downarrow
      & \hspace{-0.8cm}\downarrow
      & \hspace{-0.7cm}\downarrow
      & \hspace{-0.7cm}\downarrow
      & \hspace{-0.4cm}\downarrow \\
    \hspace{-0.0cm}    \Hgroup_p     (\RRR^t)       \hspace{-0.05cm} \to
      & \hspace{-0.35cm}\Hgroup_p     (\RRR)        \hspace{-0.05cm} \to
      & \hspace{-0.35cm}\Hgroup_p     (\RRR, \RRR^t)\hspace{-0.05cm} \to
      & \hspace{-0.35cm}\Hgroup_{p-1} (\RRR^t)      \hspace{-0.05cm} \to
      & \hspace{-0.35cm}\Hgroup_{p-1} (\RRR).
  \end{array}
 $$
 By Lemma C, the vertical maps between the first, second,
 fourth, and fifth groups are isomorphisms.
 The diagram commutes because all maps are induced by inclusion.
 The Steenrod Five Lemma thus implies that the vertical maps
 between the middle groups is also an isomorphism \cite[p.\ 140]{Mun84}.
\eop

Similar to Lemma C, Lemma D extends to the mono-chromatic case.

%%%%%%%%%%%%%%%%%%%%%%%%%%%%%%%%%%%%%%%%%%%%%%%%%%%%%%%%%%%%%%%%
\subsection{Persistent Homology}
\label{sec32}
%%%%%%%%%%%%%%%%%%%%%%%%%%%%%%%%%%%%%%%%%%%%%%%%%%%%%%%%%%%%%%%%

It is instructive to display the information contained in a persistence module
as a finite multi-set of points (referred to as \emph{dots}) in the plane.
After explaining how this is done, we prove that the
sequences of homotopy equivalent spaces introduced above
give the same diagram.

\begin{figure}[b]
 \vspace{0.0in}
 \centering
 \resizebox{!}{1.5in}{\input{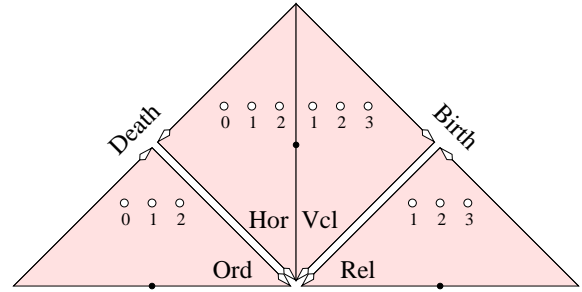}}
 \caption{From left to right: the ordinary, the horizontal, the vertical,
   and the relative subdiagram of the persistence diagram.
   The types of dots that can exist in the diagram of a 
   medusa in $4$-dimensional space-time are shown.}
 \label{fig:diagram}
\end{figure}

\paragraph{Persistence diagram.}
As before, we consider the restricted Voronoi medusa, $\RRR$,
the time function $f|_\RRR: \RRR \to [0,1]$,
and the extended module defined by the sub- and superlevel sets.
The \emph{persistence diagram} of $f|_\RRR$,
which we denote as $\Ddgm (f|_\RRR)$,
is a multi-set of dots in a double covering of the plane;
see Figure \ref{fig:diagram}.
Each dot has two coordinates
and represents a coset of homology classes.
These classes have a homological dimension, which we use to
label the dot.
Very often, we consider subdiagrams by limiting ourselves to dots
of a particular dimension.
For example, $\Ddgm_0 (f|_\RRR)$ is the multiset of dots
of homological dimension $0$, which therefore only represents components.
The arrows in our drawing indicate the direction of increasing time.
Since we have two phases, one sweeping forward and the other
backward in time, we get every pair of coordinates twice.
To explain this in more detail, we call the two coordinates
the \emph{birth} and the \emph{death} of the dot
(or of the homology classes it represents).
The birth axes appear as $-45^\circ$ lines in the figure,
and the death axes as $45^\circ$ lines.
We decompose the persistence diagram into four triangles,
referred to as the
\emph{ordinary}, \emph{horizontal}, \emph{vertical},
\emph{relative subdiagrams}, denoted as
$\Odgm (f|_\RRR)$, $\Hdgm (f|_\RRR)$, $\Vdgm (f|_\RRR)$, $\Rdgm (f|_\RRR)$.
A dot in $\Odgm (f|_\RRR)$ represents classes that are born and die
during the first phase,
a dot in $\Hdgm (f|_\RRR)$ or $\Vdgm (f|_\RRR)$ represents classes that
are born during the first phase and die during the second phase,
and a dot in $\Rdgm (f|_\RRR)$ represents classes that are born and die
during the second phase.
The difference between a dot in $\Hdgm (f|_\RRR)$ and in $\Vdgm (f|_\RRR)$
is that for the former the birth coordinate is smaller than the
death coordinate, while for the latter it is the other way round.
We believe this difference is important, as we will explain later.

\paragraph{Equivalence.}
Using Lemmas A to D, we can now make sure that the diagrams we get
from the different medusas are the same.
As before, we limit the discussion to the multi-chromatic
restricted Voronoi and alpha medusas.
Let $f|_\RRR$ and $f|_\RRR$ be the corresponding time functions.
\begin{result}[Lemma E]
  $\Ddgm(f|_\AAA)$ and $\Ddgm(f|_\RRR)$ are the same.
\end{result}
\proof
 The main tool in this proof is the diagram that connects the
 homology groups in the two modules:
 $$
  \begin{array}{cccccc}
    \hspace{-0.00cm} \Hgroup (\AAA_0)      \hspace{-0.05cm} \to
  & \hspace{-0.30cm} \ldots                \hspace{-0.05cm} \to
  & \hspace{-0.35cm} \Hgroup (\AAA_1)      \hspace{-0.05cm} \to
  & \hspace{-0.35cm} \Hgroup (\AAA,\AAA^1) \hspace{-0.05cm} \to
  & \hspace{-0.30cm} \ldots                \hspace{-0.05cm} \to
  & \hspace{-0.35cm} \Hgroup (\AAA,\AAA^0)                  \\
    \hspace{-0.45cm} \downarrow
  & 
  & \hspace{-0.80cm} \downarrow         
  & \hspace{-0.80cm} \downarrow
  & 
  & \hspace{-0.50cm} \downarrow                             \\
    \hspace{-0.00cm} \Hgroup (\RRR_0)      \hspace{-0.05cm} \to
  & \hspace{-0.30cm} \ldots                \hspace{-0.05cm} \to
  & \hspace{-0.35cm} \Hgroup (\RRR_1)      \hspace{-0.05cm} \to
  & \hspace{-0.35cm} \Hgroup (\RRR,\RRR^1) \hspace{-0.05cm} \to
  & \hspace{-0.30cm} \ldots                \hspace{-0.05cm} \to
  & \hspace{-0.35cm} \Hgroup (\RRR,\RRR^0) .
  \end{array}
 $$
 By Lemmas C and D, the vertical maps are isomorphisms.
 Since the diagram commutes,
 the claim follows by the Persistence Equivalence Theorem
 in \cite[p.\ 159]{EdHa10}.
\eop

Similar to Lemmas C and D, Lemma E extends to the mono-chromatic case,
and it implies the same relation
for Delaunay and Voronoi medusas as a special case.

%%%%%%%%%%%%%%%%%%%%%%%%%%%%%%%%%%%%%%%%%%%%%%%%%%%%%%%%%%%%%%%%
\subsection{Images}
\label{sec33}
%%%%%%%%%%%%%%%%%%%%%%%%%%%%%%%%%%%%%%%%%%%%%%%%%%%%%%%%%%%%%%%%

We expect the measurements for the alpha medusa to be more meaningful
than for the Delaunay medusa,
but there is additional information in the difference.
We compare by studying the images of the homomorphisms
induced by the inclusion of the one in the other medusa.
While we use this ability for mono-chromatic medusas,
we simplify the notation by describing it in the multi-chromatic case.

\paragraph{Image persistence.}
Recall that $\RRR_t$, $\RRR^t$, $\VVV_t$, and $\VVV^t$ are the
sublevel and superlevel sets of the Voronoi medusas, for $t \in [0,1]$.
By construction, we have $\RRR_t \subseteq \VVV_t$
and $\RRR^t \subseteq \VVV^t$.
Applying the homology functor, we get finitely many distinct groups,
which we denote as $\Rgroup_i$ for the restricted
and as $\Vgroup_i$ for the unrestricted Voronoi medusa.
Aligning the two modules, we repeat groups as necessary
and arrange them in a commuting diagram:
$$
  \begin{array}{ccccccc}
    \Rgroup_0 & \to & \Rgroup_1 & \to & \ldots & \to & \Rgroup_N  \\
    \downarrow&     & \downarrow&     &        &     & \downarrow \\
    \Vgroup_0 & \to & \Vgroup_1 & \to & \ldots & \to & \Vgroup_N .
  \end{array}
$$
Writing $\iota_i : \Rgroup_i \to \Vgroup_i$ for the homomorphism
induced by the inclusion, we are interested in the sequence of images:
\begin{eqnarray}
  0 ~=~ \image{\iota_0} \to \image{\iota_1} \to \ldots
                        \to \image{\iota_N} ~=~ 0 .
  \label{eqn:images}
\end{eqnarray}
Similar to the module of homology groups, \eqref{eqn:images}
is a sequence of vector spaces connected by homomorphisms.
We can therefore define births and deaths.
We refer to the corresponding multiset of dots
as the \emph{image persistence diagram},
denoted as $\Ddgm (\image f|_\RRR \to f|_\VVV)$;
see \cite{CEHM09} for a detailed discussion of this construction
and for an algorithm.

What does the diagram measure?
In the assumed case of the multi-chromatic restricted
included in the multi-chromatic unrestricted Voronoi medusa,
it measures nothing interesting,
simply because the groups $\Vgroup_i$ are not interesting.
This is different in the mono-chromatic case.
Here, we may have a cycle defined by data points of color $\ell$
surrounding points of color different from $\ell$.
In this case, we have a non-trivial class in a (closed or open)
sublevel set of the restricted Voronoi medusa whose image in the 
corresponding sublevel set of the unrestricted Voronoi medusa
is still non-trivial.
Indeed, a cycle in $\RRR_\ell$ gives rise to a dot in
$\Ddgm (\image f|_{\RRR_\ell} \to f|_{\VVV_\ell})$ iff it corresponds
to a hole formed by points of color different from $\ell$.
If the hole is not formed by such points, then it does not exist
in the unrestricted Voronoi medusa, the class maps to $0$,
and there is no corresponding dot in the image persistence diagram.

Instead of $\RRR_\ell \subseteq \VVV_\ell$, we can use the
inclusion of $\RRR_\ell$ in $\RRR$ to recognize when holes are
caused by interactions between different colors.
The algebraic set-up is the same, so we do not need to repeat it.
As described in the experimental Section \ref{sec5},
the latter inclusion seems to be more effective than
the inclusion in the mono-chromatic Voronoi medusa.
This is perhaps related to the fact that the most
interesting is also the most difficult case in this respect,
namely that of a $1$-dimensional homology class in $\Rspace^3$.

\paragraph{Equivalence.}
Similar to the conventional case, the image persistence
diagrams do not depend on the representation of the space and
the time function we use.
Specifically, we get the same diagrams for the inclusion of
$\RRR$ in $\VVV$ as for the inclusion of $\AAA$ in $\DDD$.
\begin{result}[Lemma F]
  $\Ddgm (\image f|_\RRR \to f|_\VVV)$ and
  $\Ddgm (\image f|_\AAA \to f|_\DDD)$ are equal as diagrams.
\end{result}
\proof
 Arrange the four modules in a $3$-dimensional diagram,
 in which the $2$-dimensional section at time $t$ consists
 of the (absolute) homology groups of the sublevel sets:
 $$
  \begin{array}{ccc}
    \Hgroup (\AAA_t) & \to & \Hgroup (\RRR_t)  \\
    \downarrow       &     & \downarrow        \\
    \Hgroup (\DDD_t) & \to & \Hgroup (\VVV_t) ,
  \end{array}
 $$
 or of the relative homology groups if the section is taken
 during the second phase.
 In all three directions, the maps are induced by inclusion,
 so the diagram commutes.
 This implies that we have maps between the corresponding 
 two modules of images.
 By Lemmas C and D, the horizontal maps in the $3$-dimensional
 diagram are isomorphisms, which implies that the maps
 connecting the two modules of images are isomorphisms.
 The claimed relation is implied by the Persistence Equivalence
 Theorem.
\eop

Similar to Lemmas C, D, and E, Lemma F extends to the mono-chromatic case.
We note that the two derived persistence diagrams are best computed
using the complexes of simplices, $\Att$ and $\Dtt$.
Indeed, these can be connected by a mapping cylinder,
giving a complex of simplices and prisms,
not unlike but different from the alpha medusa.
From this complex, we get the image persistence
diagrams by standard matrix reduction; see \cite{CEHM09}.

%%%%%%%%%%%%%%%%%%%%%%%%%%%%%%%%%%%%%%%%%%%%%%%%%%%%%%%%%%%%%%%%
%%%%%%%%%%%%%%%%%%%%%%%%%%%%%%%%%%%%%%%%%%%%%%%%%%%%%%%%%%%%%%%%
\section{Classification}
\label{sec4}
%%%%%%%%%%%%%%%%%%%%%%%%%%%%%%%%%%%%%%%%%%%%%%%%%%%%%%%%%%%%%%%%
%%%%%%%%%%%%%%%%%%%%%%%%%%%%%%%%%%%%%%%%%%%%%%%%%%%%%%%%%%%%%%%%

In this section, we discuss the information contained in the extended
persistence module defined by the time function,
interpreting the corresponding events in static space-time
as well as in dynamic temporal language.

%%%%%%%%%%%%%%%%%%%%%%%%%%%%%%%%%%%%%%%%%%%%%%%%%%%%%%%%%%%%%%%%
\subsection{Plane $\times$ Time}
\label{sec41}
%%%%%%%%%%%%%%%%%%%%%%%%%%%%%%%%%%%%%%%%%%%%%%%%%%%%%%%%%%%%%%%%

We begin with the $2$-dimensional case as a warm-up exercise,
but also to facilitate the comparison with the $3$-dimensional case.
Here, the medusa is embedded in $\Rspace^2 \times [0,1]$,
and the time function, denoted as $f$,
maps each point to its time coordinate.
We can have dots in the diagram for dimensions $0$, $1$, $2$,
except for $\Odgm_2(f)$, $\Hdgm_2(f)$, $\Vdgm_0(f)$, and $\Rdgm_0(f)$.
Recall that the persistence of the class represented by
a dot $X = (a,b)$ is the absolute difference between the
two coordinates, which we denote as $\pers{X} = |a-b|$.
In the ordinary and relative subdiagrams, $\pers{X}$ is
$\sqrt{2}$ times the distance from the baseline,
while in the horizontal and vertical subdiagrams, it is
$\sqrt{2}$ times the distance from the vertical axis that goes
through the middle of the diagram.
As established in Section \ref{sec3}, we have the choice between
several filtrations, all giving the same persistence diagram.
To focus the discussion, we limit ourselves to the medusa that
best approximates the reality of biological cells,
namely the mono-chromatic restricted Voronoi medusa.
In the text below, we use \emph{cell} in the biological
meaning of the word.
\begin{figure*}[hbt]
 \vspace{0.0in}
 \centering
 \subfigure[Ordinary subdiagram.
    Left:  a gap appears when a new component is born
    and disappears when this component merges with another component. 
    Right:  a ring appears when two ends of a component meet
    and disappears when the surrounded hole fills up.]{
   \includegraphics[width=8.1cm]{Figs/cases-ord.eps}
 } \hspace{0.7cm}
 \subfigure[Horizontal subdiagram.
    Left:  a gap appear and disappears when a new components
    is formed and ends.
    Right: a ring appears when two ends of a component meet
    and disappears when the ring breaks open.]{
   \includegraphics[width=8.1cm]{Figs/cases-hor.eps}
 }
 \subfigure[Vertical subdiagram.
    Left:  a gap appears when a component splits in two
    and disappears when the two merge back together into a single component.
    Right:  a ring appears when a disk is punctured,
    which creates a hole that first grows and later fills up.]{
   \includegraphics[width=8.1cm]{Figs/cases-vcl.eps}
 } \hspace{0.7cm}
 \subfigure[Relative subdiagram.
    Left:  a gap appears when a new component is split off
    and disappears when the component ends.
    Right:  a ring appears when the disk is punctured
    and disappears when the ring breaks open.]{
   \includegraphics[width=8.1cm]{Figs/cases-rel.eps}
 }
 \caption{The classes recorded in the four subdiagrams of the persistence diagram.}
 \label{fig:cases}
\end{figure*}

\paragraph{Type analysis.}
The classes recorded in the ordinary subdiagram are illustrated
in Figure \ref{fig:cases} (a).
A dot $X = (a,b)$ in $\Odgm_0 (f)$ corresponds to a component
born at $t = a$ that merges with another, older component at $t = b$.
If $a = 0$, then a non-empty subset of the cells that make up the
component exist at the very beginning of the observed time-series.
Its death marks the merger with another component also born at $0$,
and the tie in the decision who dies and who lives is broken arbitrarily.
If $a > 0$, the component is born because of a new cell
that is either created by cell division or enters the observation for
technical reasons.
A dot $X = (a,b)$ in $\Odgm_1 (f)$ corresponds to a ring
(an annulus) that forms at $t = a$ and whose hole fills up at $t = b$.

The classes in the horizontal subdiagram are illustrated
in Figure \ref{fig:cases} (b).
A dot $X = (a,b)$ in $\Hdgm_0 (f)$ corresponds to a component
born on the way up at $t = a$ and dying on the way down at $t = b$.
We have $a < b$ since $a$ is the minimum and $b$ is the maximum
time value of the points in the component.
Assuming some cells of color $\ell$ live from the beginning to the
end of the observations, we have at least one dot with coordinates
$0$ and $1$ in $\Hdgm_0 (f)$.
In the case of a successful segregation, in which color $\ell$
eventually forms a single component, we have exactly one such dot
and no other other dots with second coordinate equal to $1$
in the $0$-dimensional horizontal subdiagram.
A dot $X = (a,b)$ in $\Hdgm_1 (f)$ corresponds to a ring that forms at
$t = a$ and breaks up at $t = b$.
The breaking up of the ring is not detected by the homology
of the growing sublevel set, 
but rather later by the relative homology during the down phase.
The classes recorded in the vertical subdiagram are illustrated
in Figure \ref{fig:cases} (c).
A dot $X = (b,a)$ in $\Vdgm_1 (f)$ corresponds to a tunnel born on 
the way up at $t = b$ which dies on the way down at $t = a$,
where $a < b$.
The tunnel is the plane-time expression of a gap that
temporarily opens up between two subpopulations of a component
that later closes again.
A dot $X = (b,a)$ in $\Vdgm_2 (f)$ corresponds to a void born on the way
up at $t = b$ which dies on the way down at $t = a$, where $a < b$.
The void is formed by a temporary ring structure we see within an
interval of level sets.
In contrast to the rings discussed earlier,
this ring is formed by puncturing.
First, the hole expands until it eventually shrinks back
to a point and disappears at $ t = b$.

The classes recorded in the relative subdiagram are illustrated
in Figure \ref{fig:cases} (d).
A dot $X = (b,a)$ in $\Rdgm_1 (f)$ corresponds to a component
of the level set that splits off another component at $t = a$,
and dies off at $t = b$.
Since this event is detected in the down phase of the module,
we see the birth of a $1$-dimensional class at $t = b$ and
its death at $t = a$.
A dot $X = (b,a)$ in $\Rdgm_2 (f)$ corresponds to ring formed
by puncturing a disk at $t = a$.
The hole expands and eventually disappeared because of the
breaking up of the ring at $t = b$.

\paragraph{Interaction between colors.}
The appearance and disappearance of a hole may or may not
be facilitated by interactions with cells of different color.
For example, a gap recorded in the ordinary subdiagram may disappear
by squeezing out cells of color different from $\ell$, or simply
by locally consolidating the configuration.
In the former case, it is likely that the gap
also existed in $\VVV_\ell$ and unlikely that it existed in $\RRR$.
Accordingly, the gap in $\RRR_\ell$ would be represented by a dot
in the image persistence diagram of $\RRR_\ell \subseteq \VVV_\ell$
but not in that of $\RRR_\ell \subseteq \RRR$.
In the latter case, it would of course be the other way round.
Similar distinctions can be made for gaps recorded
in $\Hdgm_0$, $\Vdgm_1$, and $\Rdgm_1$.
The analysis of rings formed by cells of color $\ell$ is similar.
A ring may form around a population of cells of color different from $\ell$,
or around empty space.
In the former case, the ring can only disappear by killing the surrounded cells
or by breaking up to release these cells.
Since killing seems unlikely, this may mean that such rings are rare
in the ordinary and more common in the horizontal subdiagram.
Similarly, the latter case is likely to be the preferred configuration
for rings recorded in the vertical and the relative subdiagrams.
Indeed, these rings form by puncturing and can therefore not be invaded
by cells from the outside.
Note, however, that the suggestion that a dot in the persistence diagram
of $\RRR_\ell$ either appears or does not appear in the image persistence diagram
is an over-simplification.
Thinking of a dot as an interval, it can break up into shorter intervals,
or merge with others into a longer interval.
Similarly, there is generally no simple relationship between the
image persistence diagram and the diagram of the containing medusa.

%%%%%%%%%%%%%%%%%%%%%%%%%%%%%%%%%%%%%%%%%%%%%%%%%%%%%%%%%%%%%%%%
\subsection{Space $\times$ Time}
\label{sec42}
%%%%%%%%%%%%%%%%%%%%%%%%%%%%%%%%%%%%%%%%%%%%%%%%%%%%%%%%%%%%%%%%

Similar to the planar case, we discuss the time function
on a mono-chromatic restricted Voronoi medusa,
$f: \RRR_\ell \to [0,1]$
with $\RRR_\ell \subseteq \Rspace^3 \times [0,1]$.

\paragraph{Types.}
Compared to the planar case, we have one more dimension and therefore
one additional case in each subdiagram; see Figure \ref{fig:diagram}.
In temporal language, we follow the evolution of $0$-, $1$-, and
$2$-dimensional holes, which we refer to as
\emph{gaps}, \emph{tunnels}, and \emph{voids}.
There are two ways a hole can appear:
by \emph{closing} the surrounding cycle, or by \emph{puncturing}.
 Of course, a hole can already exist at the beginning, at $t = 0.0$,
 in which case its birth coordinate is $0.0$.
Similarly, there are two ways a hole can disappear:
by \emph{breaking up} the surrounding cycle, or by \emph{filling up} the hole.
 In addition, it may remain to the end, at $t = 1.0$,
 in which case the death coordinate is $1.0$.
We observe that a $p$-dimensional hole in the level set gives rise
to either a $p$-dimensional or a $(p+1)$-dimensional class in the persistence
module, and which it is depends on how the hole comes into existence:
either by closing a $p$-dimensional cycle or
by puncturing a $(p+1)$-dimensional chain; see Table \ref{tbl:holes}.
A hole recorded in the ordinary subdiagram is formed by closing a cycle,
allowing for the case that it exists already at the beginning,
 and it disappears by filling up.
We refer to this action as \emph{aggregation}.
Symmetrically,
 a hole recorded in the relative subdiagram is formed by puncturing,
 and the surrounding cycle either breaks up or the hole remains until the end.
We refer to this action as \emph{disaggregation}.
 In contrast, the holes recorded in the vertical and horizontal
 subdiagrams represent aggregations and disaggregations that
 are either transient or incomplete.
Specifically, a hole recorded in the vertical subdiagram appears by puncturing,
 but instead of breaking up, it disappears by filling.
Finally, a hole in the horizontal subdiagram
 forms by closing a cycle that breaks up later,
 which allows for the case that the hole exists already at the beginning
 or that is remains to the end.

\begin{table}[hbt]
  \vspace*{0.1in} \centering \small
  \begin{tabular}{c|cccc}
         & $\Odgm$    & $\Hdgm$     & $\Vdgm$       & $\Rdgm$  \\
      &~~close/fill~~ & close/break  &~~punct/fill~~ & punct/break \\ \hline
    $0$  &   gap      &   gap       &               &          \\
    $1$  &  tunnel    &  tunnel     &   gap         &   gap    \\
    $2$  &   void     &   void      &  tunnel       &  tunnel  \\
    $3$  &            &             &   void        &   void
  \end{tabular}
  \caption{The twelve cases that arise from four kinds of evolutions
    possible for each of three different types of holes in a
    $1$-parameter family of spaces in $\Rspace^3$.}
  \label{tbl:holes}
\end{table}

\begin{figure}[hbt]
 \vspace{0.0in}
 \centering
 \resizebox{!}{3.0in}{\input{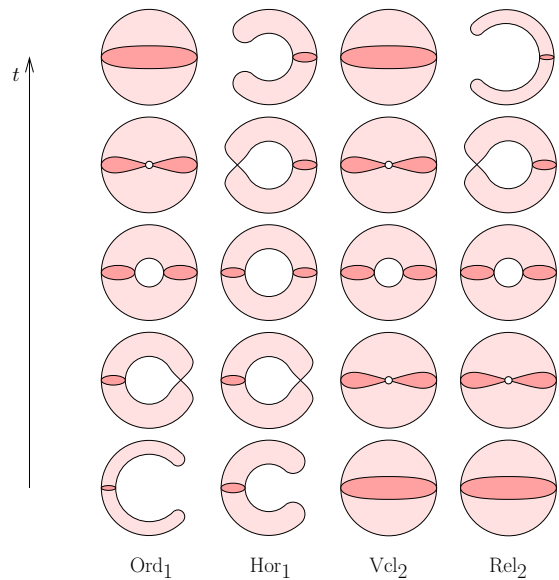}}
 \caption{From left to right:
   the evolution of a ring/tunnel with recording dot in the
   ordinary, horizontal, vertical, relative subdiagram of $\Ddgm (f)$.
   The critical configurations in the second and fourth rows separate
   the regular configurations in the other rows.}
 \label{fig:tunnels}
\end{figure}

\paragraph{Rings and tunnels.}
The $0$- and $2$-dimensional holes in $\Rspace^3$ behave like the
$0$- and $1$-dimensional holes in $\Rspace^2$, while the tunnels
introduce behavior that cannot be observed in the planar case.
Drawing the ring passing around a tunnel as a solid torus,
Figure \ref{fig:tunnels} shows the four different evolutions
recorded by dots in the four subdiagrams of $\Ddgm (f)$.

As in the other cases, it is interesting to distinguish between
tunnels formed by interactions with cells of color different from $\ell$,
from tunnels formed without such interactions.
As before, we use the image persistence diagrams induced by
$\RRR_\ell \subseteq \VVV_\ell$ or $\RRR_\ell \subseteq \RRR$.
If the tunnel corresponds to a dot in the diagram induced
by the inclusion in $\VVV_\ell$, then this is evidence for a
different color cell squeezing through an opening in the wall
of color $\ell$.
Similar evidence for this event is provided by the absence
of this dot in the diagram induced by the inclusion in $\RRR$.
The different color cell may be successful and move from one side
of the wall to the other, or it may be repelled by the force holding
the wall together and bounce back.
Which case it is may sometimes be apparent but can generally not be
read off the diagrams.

%\clearpage
%%%%%%%%%%%%%%%%%%%%%%%%%%%%%%%%%%%%%%%%%%%%%%%%%%%%%%%%%%%%%%%%%%%%%%%%%%
%%%%%%%%%%%%%%%%%%%%%%%%%%%%%%%%%%%%%%%%%%%%%%%%%%%%%%%%%%%%%%%%%%%%%%%%%%
\section{Experiments}
\label{sec5}
%%%%%%%%%%%%%%%%%%%%%%%%%%%%%%%%%%%%%%%%%%%%%%%%%%%%%%%%%%%%%%%%%%%%%%%%%%
%%%%%%%%%%%%%%%%%%%%%%%%%%%%%%%%%%%%%%%%%%%%%%%%%%%%%%%%%%%%%%%%%%%%%%%%%%

This section presents results for datasets generated with a software
that simulates the spatial motion of cells.
It serves as an illustration of our topological methods,
and as a proof of concept.
The results encourage us to proceed to the
original goal of applying the method to cells of developing
zebrafish embryos in the near future.

\paragraph{Cell segregation.}
Cell sorting involves both the segregation of a mixed population of cells
with different properties into distinct domains, 
and the active maintenance of their segregated state. 
It has been described to occur in vivo in a wide variety
of biological processes, 
such as the early embryonic development of multi-cellular organisms. 
The segregation behavior can be  studied in vitro --
in artificial cell culture conditions --
by mixing cells with different properties, 
and observing their subsequent sorting behavior. 
Such studies have been instrumental 
in revealing the cellular properties driving the segregation process. 
Different theories have been postulated and experimentally
tested to explain the mechanism that drive cell sorting
both in vitro and in vivo. 
However, to determine relative contributions of cellular properties requires 
the accurate recording as well as a quantitative and descriptive analysis 
of the sorting process in space and time.
This is the purpose of this paper,
namely a refined measurement of the spatial process that will
allow us to compare different experimental conditions
and quantify similarities and differences to wild type processes.

Here we use the zebrafish gastrulation as a case-study,
in which mesoderm cells are induced and segregate from the
overlying ectoderm germ layer. 
This process can be recapitulated in vitro by randomly mixing population 
of ectoderm and mesoderm cells in culture,
and using advanced time-lapse microscopy imaging techniques \cite{KKGH10}
to observe how mesoderm cells engulf the ectoderm cells.

\paragraph{Simulations.}
Because of the difficulties of accurately tracking the trajectories
of cell nuclei in real data, we concentrate on simulated data
to demonstrate our methods.
We use the publicly available simulation
software \emph{CompuCell3D}\footnote{http://www.compucell3d.org/}
\cite{CompuCellHandbook}
to get data sets that imitate the cell segregation process.
The software uses a $3$-dimensional version of the widely applied
\emph{Cellular Potts model} (also known as \emph{Glazier and Graner model} \cite{GG92})
to describe in vitro segregation of ectoderm and mesoderm cells.
Each cell is represented by a set of voxels (unit integer cubes in $\Rspace^3$),
and randomly tries to extend into the empty space or a neighboring cell.
An \emph{elementary change} is the alteration of the membership status of a single voxel.
The decision whether or not to accept such a change depends on
an energy function that encodes the characteristics of the dynamic process.
Evaluating the energy before and after the change,
the difference is translated into the probability of acceptance.
It is higher for negative differences, which drive the configuration
toward smaller energy.

We give more details on our specialization 
and refer to \cite{CompuCellHandbook} for the general simulation framework.
The first term of our energy function constrains the shape of the cells.
Setting the target volume to $49.0$, each cell increases the
energy by the square of its deviation from the target value.
The same holds for the surface area, where we set the target to $26.0$.
The software allows for weights to control the relative influence of these
terms, and we use weights $100.0$ for the volume and $3.0$ for the area.
The second term of our energy function influences the inter-cellular behavior.
We call two neighboring voxels an \emph{interface} if they
belongs to different cells, or to a cell and empty space.
In our setup, an interface between two cells decreases the energy,
so that cells tend to stick to each other.
Specifically, we define two cell types, which we call \emph{red}
(simulating ectoderm) and \emph{blue} (simulating mesoderm).
An interface between two blue cells, or between a red and a blue cell
decreases the energy by $200.0$,
while an interface between two red cells decreases the energy by $400.0$.
This results in red cells sticking to each other stronger than blue cells.
Finally, an interface between a cell and empty space
increases the energy by $20.0$ for a blue cell,
and by $100.0$ for a red cell, with the effect
that cells generally avoid contact with the surrounding space,
and red cells do so with more determination than blue cells.

\paragraph{Datasets.}
We construct the initial configuration of our experiments by tiling
the domain into $6\times 6\times 6 = 216$ cubes and placing one cell
into each cube to fill it completely.
This setup leads to cells that are larger than the target volume,
resulting in a high initial value of the energy function.
In the first steps of the simulation,
the cells shrink and cluster in the middle of the available space.
However, some of them lose contact and stay separated from the bulk
for a long time, if not the entire time.
\begin{figure}[h]
 \vspace{-0.1in} \centering
   \includegraphics[width=4cm]{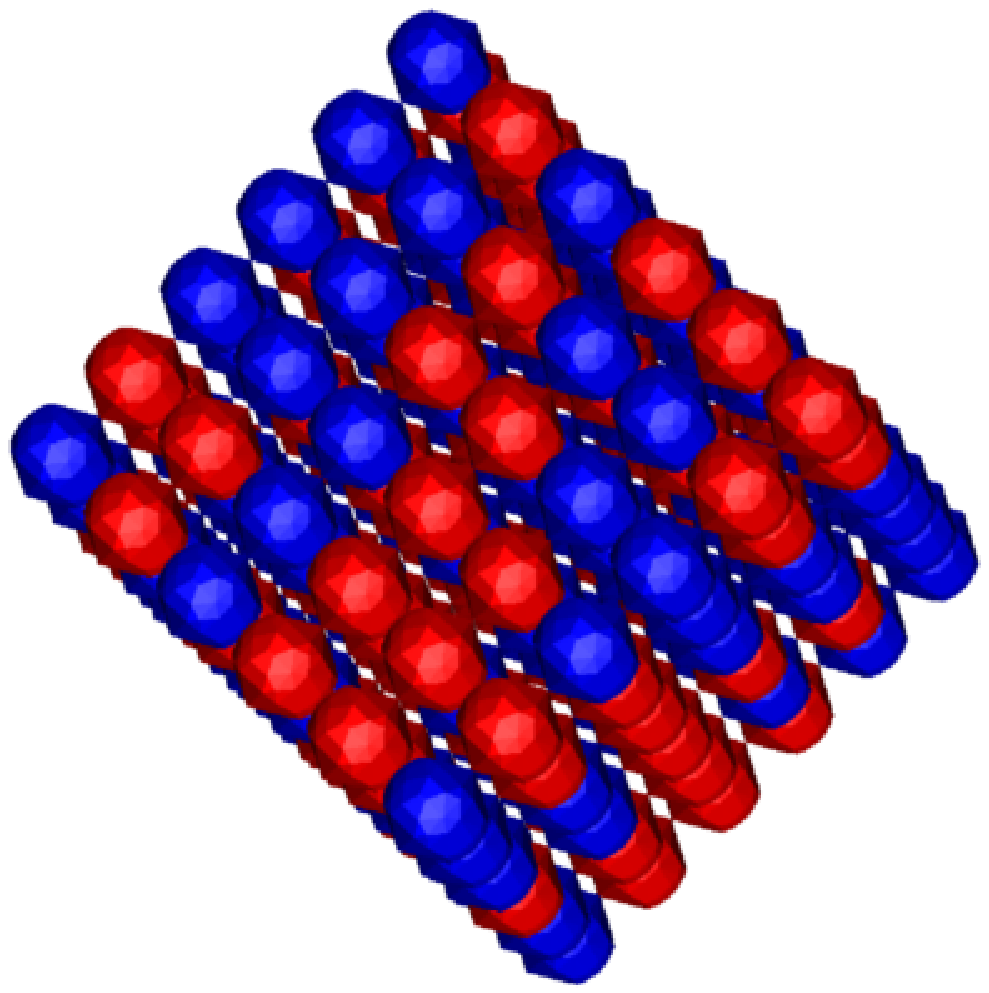}
   \includegraphics[width=4cm]{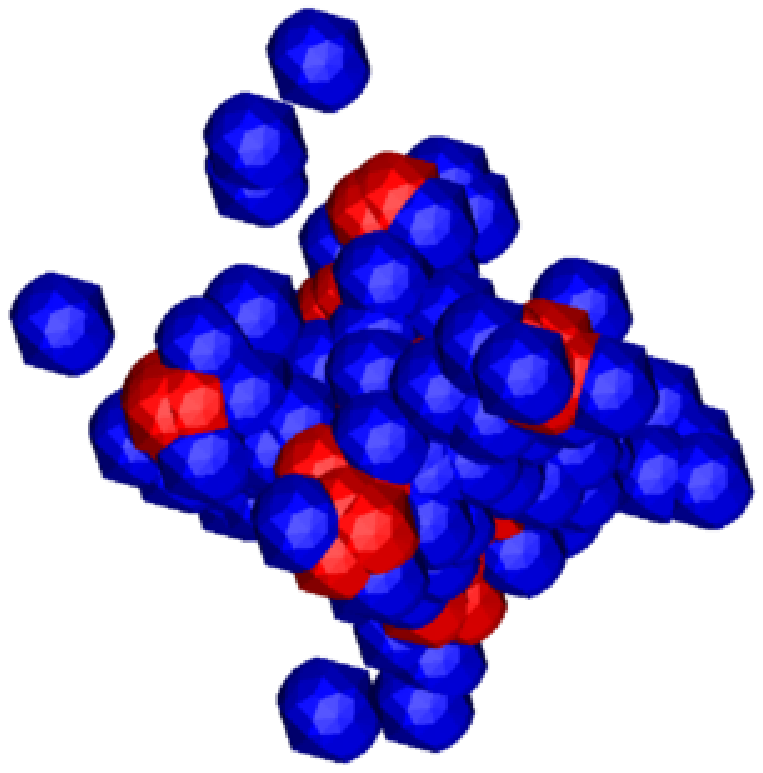}
   \includegraphics[width=4cm]{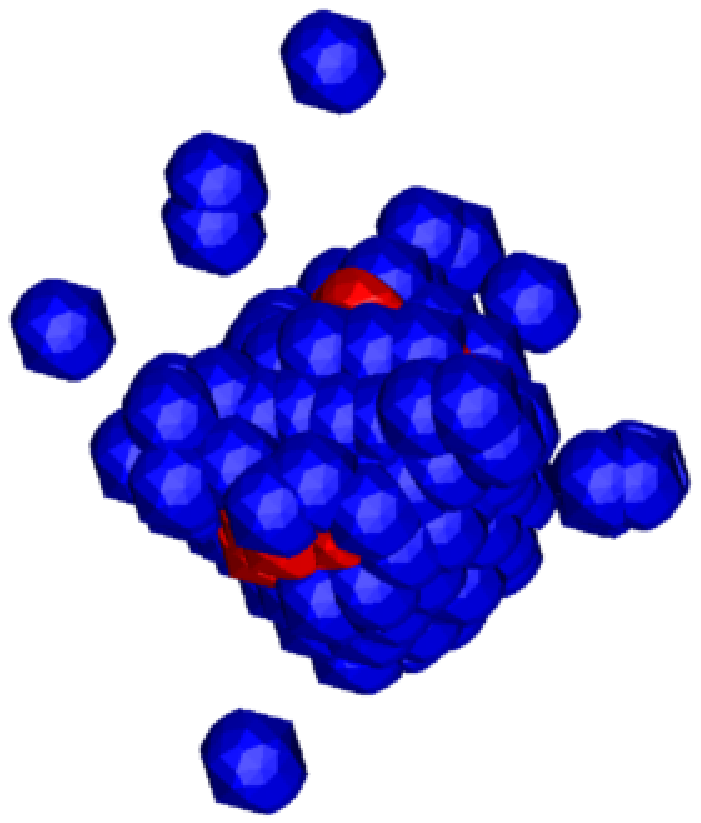}
   \includegraphics[width=4cm]{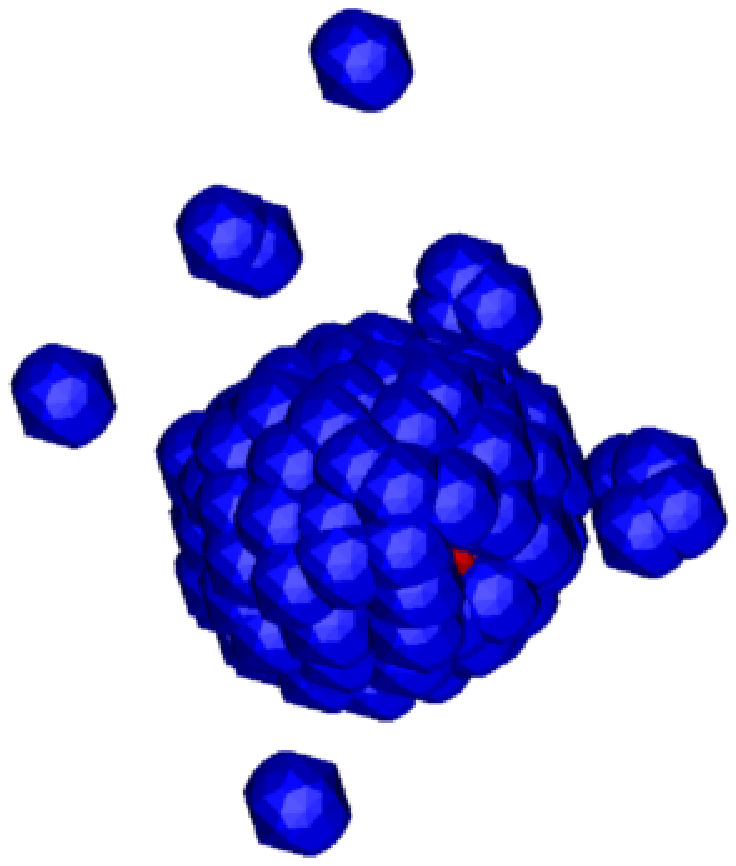}
 \vspace{-0.2in}
 \caption{The restricted Voronoi tessellation at four moments in time.
   At the beginning, the cells form a cubical grid (upper-left). 
   Next, the cells move toward the middle of the available space,
   leaving some outliers behind (upper-right).
   Thereafter, the blue cells begin to engulf the red cells (lower-left).
   Finally, the blue cells form a sphere surrounding a ball of
   red cells (lower-right).}
 \label{fig:Vor-tessel}
\end{figure}
We consider three simulations.
In the first, we color each cell either red or blue according to a fair coin flip;
see Figure \ref{fig:Vor-tessel}.
This intermixed configuration mimics the biologically interesting process of
cell segregation.
The second and third are control experiments in which all cells are
of the same type, either red or blue.
To obtain trajectories, we compute the center of gravity of each cell
at various moments in time and interpolate them piecewise linearly.
In all experiments, we set the radius of the ball
approximating one cell to $\alpha_0 = 4.0$, and we model the arrangement
of cells by the thus defined restricted Voronoi tessellation.

As explained in the preceding sections, the behavior of the cell population
in time is characterized by the persistence and image persistence diagrams
of the time function restricted to the various medusas.
There are too many diagrams to fit in this paper,
so we show a subset, selected to illustrate interesting behavior of the data.
For each diagram, we count the dots and we compute the \emph{$1$-norm},
which we define as
\begin{eqnarray}
  \norm{\Ddgm (f)}_1  &=&  \sum_{X \in \Ddgm (f)} \pers{X} .
\end{eqnarray}
These numbers are collected for each dimension and each subdiagram
separately, giving a good overall impression of the activity
recorded in the diagram.

\begin{figure}[h]
 \vspace{0.0in} \centering
   \includegraphics[width=8cm]{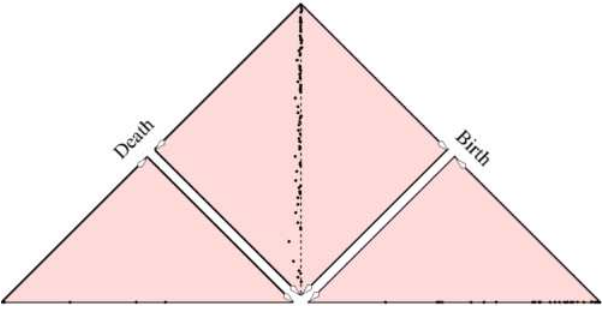}
 \caption{The $2$-dimensional persistence diagrams of
   the restricted Voronoi medusa of $\bbb$.}
 \label{fig:Multi}
\end{figure}
\paragraph{Multi-chromatic data.}
We start with the discussion of the intermixed simulation,
letting $\rrr$ and $\bbb$ denote the corresponding sets of red and blue cells.
See Figure \ref{fig:Vor-tessel} 
for illustrations of the restricted Voronoi tessellation,
Figure \ref{fig:Multi} for one of the persistence diagrams,
and Table \ref{tbl:Multi} for the quantitative summaries of all diagrams,
divided up into dimensions and subdiagrams.
The only high-persistence feature in $\rrr$ is recorded in the
horizontal subdiagram and simply represents the component itself. 
There are additional high-persistence features recorded in the
$1$-dimensional vertical and relative subdiagrams of $\bbb$,
representing the isolated (blue) cells that lose contact to the bulk
during the initial phase.
Cells recorded in the vertical subdiagram merge back into the bulk,
while cells recorded in the relative subdiagram remain separate
until the end of the simulation.
In addition, the $2$-dimensional horizontal subdiagram
of $\bbb$ records features whose persistence adds up to a value of $0.76$.
They describe the sphere of blue cells that wrap around the red cells.
The fact that it is represented by several instead of a single dot
indicates that the sphere is pinched from time to time during the simulation;
compare with Figure \ref{fig:Vor-tessel}.

To cast additional light on the process, we show results for the
blue restricted Voronoi medusa included in the blue unrestricted Voronoi medusa;
compare the two bottom panels of Table \ref{tbl:Multi}.
We no longer see any persistent features in the vertical or relative subdiagrams
because the outliers are merged with the bulk by the unrestricted medusa.
Furthermore, the image persistence diagram features a single horizontal void
with high persistence, which is the consolidation of the many
low-persistence horizontal voids of the blue restricted Voronoi medusa.
\begin{table}[h]
  \vspace{0.0in} \centering \footnotesize
  \begin{tabular}{c|rr|rr|rr|rr|rr}
   $\rrr$  & \multicolumn{2}{c|}{$\Odgm$} & \multicolumn{2}{|c|}{$\Hdgm$} & \multicolumn{2}{|c|}{$\Vdgm$} & \multicolumn{2}{|c|}{$\Rdgm$} & \multicolumn{2}{c}{sum} \\
    \hline
    gaps    & \hspace{-0.07in}  17 & \hspace{-0.1in} 0.03 \hspace{-0.07in}
            & \hspace{-0.07in}   1 & \hspace{-0.1in} 1.00 \hspace{-0.07in}
            & \hspace{-0.07in}   7 & \hspace{-0.1in} 0.03 \hspace{-0.07in}
            & \hspace{-0.07in}   0 & \hspace{-0.1in} 0.00 \hspace{-0.07in}
            & \hspace{-0.07in}  25 & \hspace{-0.1in} 1.07 \hspace{-0.07in} \\
    \hspace{-0.1in}tunnels\hspace{-0.1in}
            & \hspace{-0.07in} 103 & \hspace{-0.1in} 0.04 \hspace{-0.07in}
            & \hspace{-0.07in}   5 & \hspace{-0.1in} 0.00 \hspace{-0.07in}
            & \hspace{-0.07in}   0 & \hspace{-0.1in} 0.00 \hspace{-0.07in}
            & \hspace{-0.07in}   1 & \hspace{-0.1in} 0.00 \hspace{-0.07in}
            & \hspace{-0.07in} 109 & \hspace{-0.1in} 0.05 \hspace{-0.07in} \\
    voids   & \hspace{-0.07in}  21 & \hspace{-0.1in} 0.00 \hspace{-0.07in}
            & \hspace{-0.07in}   0 & \hspace{-0.1in} 0.00 \hspace{-0.07in}
            & \hspace{-0.07in}   2 & \hspace{-0.1in} 0.00 \hspace{-0.07in}
            & \hspace{-0.07in}   0 & \hspace{-0.1in} 0.00 \hspace{-0.07in}
            & \hspace{-0.07in}  23 & \hspace{-0.1in} 0.00 \hspace{-0.07in} \\
    \hline
    sum     & \hspace{-0.07in} 141 & \hspace{-0.1in} 0.08 \hspace{-0.07in}
            & \hspace{-0.07in}   6 & \hspace{-0.1in} 1.00 \hspace{-0.07in}
            & \hspace{-0.07in}   9 & \hspace{-0.1in} 0.03 \hspace{-0.07in}
            & \hspace{-0.07in}   1 & \hspace{-0.1in} 0.00 \hspace{-0.07in}
            & \hspace{-0.07in} 157 & \hspace{-0.1in} 1.12 \hspace{-0.07in} \\
    \hline \hline
   $\bbb$   & & & & &              & & & & &                               \\
    \hline
    gaps    & \hspace{-0.07in}  10 & \hspace{-0.1in} 0.02 \hspace{-0.07in}
            & \hspace{-0.07in}   1 & \hspace{-0.1in} 1.00 \hspace{-0.07in}
            & \hspace{-0.07in}  22 & \hspace{-0.1in} 1.58 \hspace{-0.07in}
            & \hspace{-0.07in}   3 & \hspace{-0.1in} 2.55 \hspace{-0.07in}
            & \hspace{-0.07in}  36 & \hspace{-0.1in} 5.15 \hspace{-0.07in} \\
    \hspace{-0.1in}tunnels\hspace{-0.1in}
            & \hspace{-0.07in} 237 & \hspace{-0.1in} 0.27 \hspace{-0.07in}
            & \hspace{-0.07in}  37 & \hspace{-0.1in} 0.05 \hspace{-0.07in}
         &\hspace{-0.07in}{\bf  21}&\hspace{-0.13in}{\bf 0.04}\hspace{-0.04in}
         &\hspace{-0.07in}{\bf  41}&\hspace{-0.13in}{\bf 0.02}\hspace{-0.04in}
            & \hspace{-0.07in} 336 & \hspace{-0.1in} 0.40 \hspace{-0.07in} \\
    voids&\hspace{-0.07in}{\bf  32}&\hspace{-0.13in}{\bf 0.00}\hspace{-0.04in}
         &\hspace{-0.07in}{\bf  61}&\hspace{-0.13in}{\bf 0.76}\hspace{-0.04in}
            & \hspace{-0.07in}   0 & \hspace{-0.1in} 0.00 \hspace{-0.07in}
            & \hspace{-0.07in}   0 & \hspace{-0.1in} 0.00 \hspace{-0.07in}
            & \hspace{-0.07in}  93 & \hspace{-0.1in} 0.76 \hspace{-0.07in} \\
    \hline
    sum     & \hspace{-0.07in} 279 & \hspace{-0.1in} 0.30 \hspace{-0.07in}
            & \hspace{-0.07in}  99 & \hspace{-0.1in} 1.82 \hspace{-0.07in}
            & \hspace{-0.07in}  43 & \hspace{-0.1in} 1.62 \hspace{-0.07in}
            & \hspace{-0.07in}  44 & \hspace{-0.1in} 2.57 \hspace{-0.07in}
            & \hspace{-0.07in} 465 & \hspace{-0.1in} 6.32 \hspace{-0.07in} \\
    \hline \hline
    $\bbb$, Vor & & & & &          & & & & &                               \\
    \hline
    gaps    & \hspace{-0.07in}   0 & \hspace{-0.1in} 0.00 \hspace{-0.07in}
            & \hspace{-0.07in}   1 & \hspace{-0.1in} 1.00 \hspace{-0.07in}
            & \hspace{-0.07in}   0 & \hspace{-0.1in} 0.00 \hspace{-0.07in}
            & \hspace{-0.07in}   0 & \hspace{-0.1in} 0.00 \hspace{-0.07in}
            & \hspace{-0.07in}   1 & \hspace{-0.1in} 1.00 \hspace{-0.07in} \\
    \hspace{-0.1in}tunnels\hspace{-0.1in}
            & \hspace{-0.07in}   9 & \hspace{-0.1in} 0.02 \hspace{-0.07in}
            & \hspace{-0.07in}   4 & \hspace{-0.1in} 0.01 \hspace{-0.07in}
            & \hspace{-0.07in}   3 & \hspace{-0.1in} 0.00 \hspace{-0.07in}
            & \hspace{-0.07in}   7 & \hspace{-0.1in} 0.00 \hspace{-0.07in}
            & \hspace{-0.07in}  23 & \hspace{-0.1in} 0.04 \hspace{-0.07in} \\
    voids   & \hspace{-0.07in}   1 & \hspace{-0.1in} 0.00 \hspace{-0.07in}
            & \hspace{-0.07in}   1 & \hspace{-0.1in} 0.95 \hspace{-0.07in}
            & \hspace{-0.07in}   0 & \hspace{-0.1in} 0.00 \hspace{-0.07in}
            & \hspace{-0.07in}   0 & \hspace{-0.1in} 0.00 \hspace{-0.07in}
            & \hspace{-0.07in}   2 & \hspace{-0.1in} 0.95 \hspace{-0.07in} \\
    \hline
    sum     & \hspace{-0.07in}  10 & \hspace{-0.1in} 0.02 \hspace{-0.07in}
            & \hspace{-0.07in}   6 & \hspace{-0.1in} 1.96 \hspace{-0.07in}
            & \hspace{-0.07in}   3 & \hspace{-0.1in} 0.00 \hspace{-0.07in}
            & \hspace{-0.07in}   7 & \hspace{-0.1in} 0.00 \hspace{-0.07in}
            & \hspace{-0.07in}  26 & \hspace{-0.1in} 2.00 \hspace{-0.07in}
    \end{tabular}
    \caption{From top to bottom:  the sizes and $1$-norms of the persistence diagrams
      of the red and blue restricted Voronoi medusas,
      and of the image persistence diagrams defined by the inclusion
      of the blue restricted Voronoi medusa in the blue Voronoi medusa.
      Boldface numbers correspond to dots in Figure \ref{fig:Multi}.}
    \label{tbl:Multi}
\end{table}

\begin{figure}[h]
 \vspace{0.0in} \centering
   \includegraphics[width=8cm]{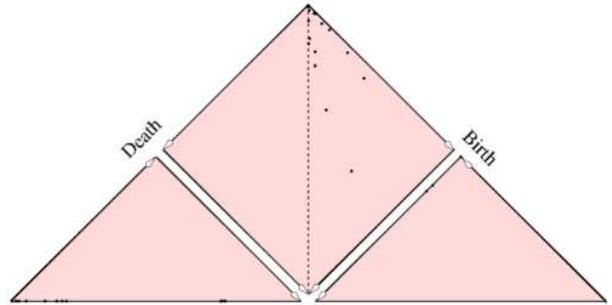}
 \caption{The $1$-dimensional persistence diagrams of the restricted
   Voronoi medusa of $\bbb\bbb$.}
 \label{fig:Mono}
\end{figure}
\paragraph{Mono-chromatic data.}
The control experiments use simulations with cells of a single type,
which is either red or blue.
To avoid confusion with the data from the previous experiment,
we denote the mono-chromatic sets of cells by $\rrr\rrr$ and $\bbb\bbb$.
Table \ref{tbl:Mono} presents the quantitative summaries of the
topological analysis.
The process looks similar for both colors:
there is a component formed by the majority of the cells
(represented by a dot in horizontal subdiagram),
and there are several outlier cells
(represented by dots in the vertical and relative subdiagram).
Moreover, many low-persistence ordinary tunnels are
created in the initial phase, 
whose persistence adds up to a significant number.

There are also differences.
First, the blue cells create many vertical gaps
(components that split and re-merge during the simulation),
while they are rare for the red cells.
An inspection of Figure \ref{fig:Mono} reveals that most of these splits
happen early in the simulation, when the cells still shrink.
This effect can be explained by the smaller stickiness of the blue cells,
which makes it more likely for them to lose contact with the bulk.
Once separated, outliers move randomly in space and find contact with the
bulk only by chance.
Curiously, the red medusa has the same number of relative gaps
(components that split and the pieces stay separate)
as the blue medusa.
It suggests that the red outliers find it more difficult to merge
back into the bulk.
One reason may be their stickiness to each other, and their phobia
of the empty space, which limits their search activity.

A second difference can be observed in the vertical voids, which only
exist for the red cells.
Recall that they are created by puncturing and destroyed by filling.
As we can read off the table, these holes are short-lived.
There is no biological reason why cells would create such holes,
and they should indeed be considered an artifact of the abstraction
of cells as relatively inflexible restricted Voronoi regions.
Nevertheless, even this artifact tells us something about the simulation process.
The presence of such holes means that the red cells tend
to deviate more form the shape of a round ball than the blue cells.
This is consistent with the energy function in our setup
in which red-red interfaces are more favorable than blue-blue interfaces.
Therefore, the red cells get more reward for increasing their interface area 
at the expense of deviating from the target volume and surface area.

\begin{table}[h]
  \vspace{0.0in} \centering \footnotesize
  \begin{tabular}{c|rr|rr|rr|rr|rr}
    $\rrr\rrr$ & \multicolumn{2}{c|}{$\Odgm$} & \multicolumn{2}{|c|}{$\Hdgm$} & \multicolumn{2}{|c|}{$\Vdgm$} & \multicolumn{2}{|c|}{$\Rdgm$} & \multicolumn{2}{c}{sum} \\
    \hline
    gaps    & \hspace{-0.07in}   0 & \hspace{-0.1in} 0.00 \hspace{-0.07in}
            & \hspace{-0.07in}   1 & \hspace{-0.1in} 1.00 \hspace{-0.07in}
            & \hspace{-0.07in}   2 & \hspace{-0.1in} 0.01 \hspace{-0.07in}
            & \hspace{-0.07in}   3 & \hspace{-0.1in} 2.99 \hspace{-0.07in}
            & \hspace{-0.07in}   6 & \hspace{-0.1in} 4.01 \hspace{-0.07in} \\
    \hspace{-0.1in}tunnels\hspace{-0.1in}
            & \hspace{-0.07in} 607 & \hspace{-0.1in} 0.60 \hspace{-0.07in}
            & \hspace{-0.07in}  47 & \hspace{-0.1in} 0.04 \hspace{-0.07in}
            & \hspace{-0.07in}   1 & \hspace{-0.1in} 0.00 \hspace{-0.07in}
            & \hspace{-0.07in}   3 & \hspace{-0.1in} 0.00 \hspace{-0.07in}
            & \hspace{-0.07in} 658 & \hspace{-0.1in} 0.64 \hspace{-0.07in} \\
    voids   & \hspace{-0.07in} 339 & \hspace{-0.1in} 0.03 \hspace{-0.07in}
            & \hspace{-0.07in}   0 & \hspace{-0.1in} 0.00 \hspace{-0.07in}
            & \hspace{-0.07in}  24 & \hspace{-0.1in} 0.02 \hspace{-0.07in}
            & \hspace{-0.07in}  10 & \hspace{-0.1in} 0.00 \hspace{-0.07in}
            & \hspace{-0.07in} 373 & \hspace{-0.1in} 0.06 \hspace{-0.07in} \\
    \hline
    sum     & \hspace{-0.07in} 946 & \hspace{-0.1in} 0.64 \hspace{-0.07in}
            & \hspace{-0.07in}  48 & \hspace{-0.1in} 1.04 \hspace{-0.07in}
            & \hspace{-0.07in}  27 & \hspace{-0.1in} 0.04 \hspace{-0.07in}
            & \hspace{-0.07in}  16 & \hspace{-0.1in} 2.99 \hspace{-0.07in}
            & \hspace{-0.07in}1037 & \hspace{-0.1in} 4.72 \hspace{-0.07in} \\
    \hline \hline
    $\bbb\bbb$ & & & & &  \\
    \hline
    gaps    & \hspace{-0.07in}   0 & \hspace{-0.1in} 0.00 \hspace{-0.07in}
            & \hspace{-0.07in}   1 & \hspace{-0.1in} 1.00 \hspace{-0.07in}
            & \hspace{-0.07in}  {\bf 15} & \hspace{-0.1in} {\bf 1.52} \hspace{-0.07in}
            & \hspace{-0.07in}  {\bf 3} & \hspace{-0.1in} {\bf 2.56} \hspace{-0.07in}
            & \hspace{-0.07in}  19 & \hspace{-0.1in} 5.08 \hspace{-0.07in} \\
    \hspace{-0.1in}tunnels\hspace{-0.1in}
            & \hspace{-0.07in} {\bf 582} & \hspace{-0.1in} {\bf 0.55} \hspace{-0.07in}
            & \hspace{-0.07in} {\bf 81} & \hspace{-0.1in} {\bf 0.09} \hspace{-0.07in}
            & \hspace{-0.07in}   0 & \hspace{-0.1in} 0.00 \hspace{-0.07in}
            & \hspace{-0.07in}  16 & \hspace{-0.1in} 0.00 \hspace{-0.07in}
            & \hspace{-0.07in} 679 & \hspace{-0.1in} 0.65 \hspace{-0.07in} \\
    voids   & \hspace{-0.07in} 310 & \hspace{-0.1in} 0.02 \hspace{-0.07in}
            & \hspace{-0.07in}   0 & \hspace{-0.1in} 0.00 \hspace{-0.07in}
            & \hspace{-0.07in}   0 & \hspace{-0.1in} 0.00 \hspace{-0.07in}
            & \hspace{-0.07in}   4 & \hspace{-0.1in} 0.00 \hspace{-0.07in}
            & \hspace{-0.07in} 314 & \hspace{-0.1in} 0.02 \hspace{-0.07in} \\
    \hline
    sum     & \hspace{-0.07in} 892 & \hspace{-0.1in} 0.58 \hspace{-0.07in}
            & \hspace{-0.07in}  82 & \hspace{-0.1in} 1.09 \hspace{-0.07in}
            & \hspace{-0.07in}  15 & \hspace{-0.1in} 1.52 \hspace{-0.07in}
            & \hspace{-0.07in}  23 & \hspace{-0.1in} 2.56 \hspace{-0.07in}
            & \hspace{-0.07in}1012 & \hspace{-0.1in} 5.76 \hspace{-0.07in} \\
    \end{tabular}
    \caption{The sizes and $1$-norms of the persistence diagrams
      of the red medusa at the top and the blue medusa at the bottom.
      Boldface numbers correspond to dots in Figure \ref{fig:Mono}.}
    \label{tbl:Mono}
\end{table}

\begin{figure}[hbt]
 \vspace{0.0in} \centering
 \subfigure[Image diagram induced by the inclusion of the restricted
   Voronoi medusa of $\bbb_1$ in unrestricted Voronoi medusa of the
   same set of cells.]{
   \includegraphics[width=8cm]{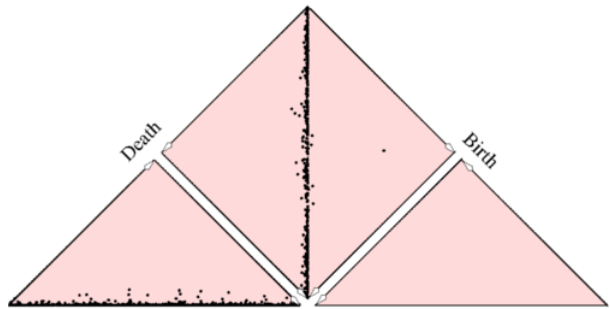}}
 \subfigure[Image diagram induced by the inclusion of the
   restricted Voronoi medusa of $\bbb_1$ in the restricted Voronoi
   medusa of $\bbb\bbb$.]{
   \includegraphics[width=8cm]{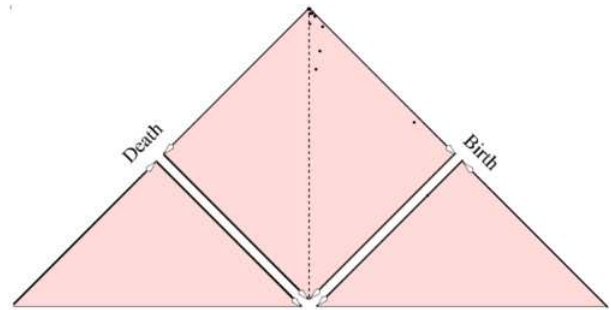}}
 \caption{The $1$-dimensional image persistence diagrams of the inclusion
   of $\bbb_1$ in its corresponding Voronoi medusa at the top,
   and the blue combined medusa at the bottom.}
 \label{fig:Blue}
\end{figure}
\paragraph{Image diagrams.}
We create a randomly intermixed process by decomposing $\bbb\bbb$
into two sets of cells, $\bbb_1$ and $\bbb_2$, with fair coin flips. 
There is no reason to expect a sorting process similar to the
simulation with red and blue cells. Instead, the two (identical) cell types
form a random pattern with arbitrary neighbor exchanges over time.

The quantitative results for $\bbb_1$ are given in Table \ref{tbl:Blue};
those for $\bbb_2$ are similar.
Note the large number of tunnels, which are distributed
over the entire simulation.
Recall that we have two possibilities for computing image persistence:
by mapping the restricted Voronoi medusa of $\bbb_1$
into the unrestricted Voronoi medusa of the same set of cells,
or into the restricted Voronoi medusa of $\bbb\bbb$.
For comparison, we display the quantitative results
of both inclusions in Table \ref{tbl:Blue}, showing the corresponding
$1$-dimensional diagrams in Figure \ref{fig:Blue}.
The majority of features -- both in terms of number and persistence --
carry over to the unrestricted Voronoi medusa,
which suggests that most of the holes in the medusa of $\bbb_1$
are caused by the presence of cells in $\bbb_2$. 
This is consistent with the relative sparsity of the image diagrams
defined by the inclusion of $\bbb_1$ in $\bbb\bbb$.
Notable exceptions are the gaps recorded in the $1$-dimensional
vertical and relative subdiagrams,
which describe the outliers separated from the bulk mentioned earlier.
\begin{table}[h]
  \vspace{0.0in} \centering \footnotesize
  \begin{tabular}{c|rr|rr|rr|rr|rr}
   $\bbb_1$ & \multicolumn{2}{c|}{$\Odgm$} & \multicolumn{2}{|c|}{$\Hdgm$} & \multicolumn{2}{|c|}{$\Vdgm$} & \multicolumn{2}{|c|}{$\Rdgm$} & \multicolumn{2}{c}{sum} \\
    \hline
    gaps    & \hspace{-0.07in}  11 & \hspace{-0.1in} 0.00 \hspace{-0.07in}
            & \hspace{-0.07in}   1 & \hspace{-0.1in} 1.00 \hspace{-0.07in}
            & \hspace{-0.07in}  42 & \hspace{-0.1in} 1.28 \hspace{-0.07in}
            & \hspace{-0.07in}   1 & \hspace{-0.1in} 0.76 \hspace{-0.07in}
            & \hspace{-0.07in}  55 & \hspace{-0.1in} 3.05 \hspace{-0.07in} \\
    \hspace{-0.1in}tunnels\hspace{-0.1in}
            & \hspace{-0.07in}1406 & \hspace{-0.1in} 4.32 \hspace{-0.07in}
            & \hspace{-0.07in} 629 & \hspace{-0.1in} 5.41 \hspace{-0.07in}
            & \hspace{-0.07in} 648 & \hspace{-0.1in} 6.02 \hspace{-0.07in}
            & \hspace{-0.07in}1273 & \hspace{-0.1in} 4.60 \hspace{-0.07in}
            & \hspace{-0.07in}3956 & \hspace{-0.1in}20.36 \hspace{-0.07in} \\
    voids   & \hspace{-0.07in}  32 & \hspace{-0.1in} 0.00 \hspace{-0.07in}
            & \hspace{-0.07in}   0 & \hspace{-0.1in} 0.00 \hspace{-0.07in}
            & \hspace{-0.07in}   0 & \hspace{-0.1in} 0.00 \hspace{-0.07in}
            & \hspace{-0.07in}   0 & \hspace{-0.1in} 0.00 \hspace{-0.07in}
            & \hspace{-0.07in}  32 & \hspace{-0.1in} 0.00 \hspace{-0.07in} \\
    \hline
    sum     & \hspace{-0.07in}1449 & \hspace{-0.1in} 4.33 \hspace{-0.07in}
            & \hspace{-0.07in} 630 & \hspace{-0.1in} 6.41 \hspace{-0.07in}
            & \hspace{-0.07in} 690 & \hspace{-0.1in} 7.31 \hspace{-0.07in}
            & \hspace{-0.07in}1274 & \hspace{-0.1in} 5.36 \hspace{-0.07in}
            & \hspace{-0.07in}4043 & \hspace{-0.1in}23.42 \hspace{-0.07in} \\
    \hline \hline
   \!\!\!\!$\bbb_1$, Vor\!\!\!\! & & & & &  & & & & &                      \\
    \hline
    gaps    & \hspace{-0.07in}   0 & \hspace{-0.1in} 0.00 \hspace{-0.07in}
            & \hspace{-0.07in}   1 & \hspace{-0.1in} 1.00 \hspace{-0.07in}
         &\hspace{-0.07in}{\bf  10}&\hspace{-0.13in}{\bf 0.66}\hspace{-0.04in}
         &\hspace{-0.07in}{\bf   0}&\hspace{-0.13in}{\bf 0.00}\hspace{-0.04in}
            & \hspace{-0.07in}  11 & \hspace{-0.1in} 1.66 \hspace{-0.07in} \\
    \hspace{-0.1in}tunnels\hspace{-0.1in}
        &\hspace{-0.07in}{\bf 1113}&\hspace{-0.13in}{\bf 4.05}\hspace{-0.04in}
        &\hspace{-0.07in}{\bf  445}&\hspace{-0.13in}{\bf 3.70}\hspace{-0.04in}
            & \hspace{-0.07in} 627 & \hspace{-0.1in} 5.92 \hspace{-0.07in}
            & \hspace{-0.07in}1183 & \hspace{-0.1in} 4.47 \hspace{-0.07in}
            & \hspace{-0.07in}3368 & \hspace{-0.1in}18.15 \hspace{-0.07in} \\
    voids   & \hspace{-0.07in}   0 & \hspace{-0.1in} 0.00 \hspace{-0.07in}
            & \hspace{-0.07in}   2 & \hspace{-0.1in} 0.00 \hspace{-0.07in}
            & \hspace{-0.07in}   0 & \hspace{-0.1in} 0.00 \hspace{-0.07in}
            & \hspace{-0.07in}   0 & \hspace{-0.1in} 0.00 \hspace{-0.07in}
            & \hspace{-0.07in}   2 & \hspace{-0.1in} 0.00 \hspace{-0.07in} \\
    \hline
    sum     & \hspace{-0.07in}1113 & \hspace{-0.1in} 4.05 \hspace{-0.07in}
            & \hspace{-0.07in} 448 & \hspace{-0.1in} 4.70 \hspace{-0.07in}
            & \hspace{-0.07in} 637 & \hspace{-0.1in} 6.58 \hspace{-0.07in}
            & \hspace{-0.07in}1183 & \hspace{-0.1in} 4.47 \hspace{-0.07in}
            & \hspace{-0.07in}3381 & \hspace{-0.1in}19.82 \hspace{-0.07in} \\
    \hline \hline
    \!\!\!\!$\bbb_1$, $\bbb\bbb$\!\!\!\!  & & & & &   & & & & &           \\
    \hline
    gaps    & \hspace{-0.07in}   0 & \hspace{-0.1in} 0.00 \hspace{-0.07in}
            & \hspace{-0.07in}   1 & \hspace{-0.1in} 1.00 \hspace{-0.07in}
         &\hspace{-0.07in}{\bf  15}&\hspace{-0.13in}{\bf 1.04}\hspace{-0.04in}
         &\hspace{-0.07in}{\bf   1}&\hspace{-0.13in}{\bf 0.76}\hspace{-0.04in}
            & \hspace{-0.07in}  17 & \hspace{-0.1in} 2.80 \hspace{-0.07in} \\
    \hspace{-0.1in}tunnels\hspace{-0.1in}
         &\hspace{-0.07in}{\bf 149}&\hspace{-0.13in}{\bf 0.11}\hspace{-0.04in}
         &\hspace{-0.07in}{\bf  15}&\hspace{-0.13in}{\bf 0.01}\hspace{-0.04in}
            & \hspace{-0.07in}   0 & \hspace{-0.1in} 0.00 \hspace{-0.07in}
            & \hspace{-0.07in}   1 & \hspace{-0.1in} 0.00 \hspace{-0.07in}
            & \hspace{-0.07in} 165 & \hspace{-0.1in} 0.13 \hspace{-0.07in} \\
    voids   & \hspace{-0.07in}  32 & \hspace{-0.1in} 0.00 \hspace{-0.07in}
            & \hspace{-0.07in}   0 & \hspace{-0.1in} 0.00 \hspace{-0.07in}
            & \hspace{-0.07in}   0 & \hspace{-0.1in} 0.00 \hspace{-0.07in}
            & \hspace{-0.07in}   0 & \hspace{-0.1in} 0.00 \hspace{-0.07in}
            & \hspace{-0.07in}  32 & \hspace{-0.1in} 0.00 \hspace{-0.07in} \\
    \hline
    sum     & \hspace{-0.07in} 181 & \hspace{-0.1in} 0.11 \hspace{-0.07in}
            & \hspace{-0.07in}  16 & \hspace{-0.1in} 1.01 \hspace{-0.07in}
            & \hspace{-0.07in}  15 & \hspace{-0.1in} 1.04 \hspace{-0.07in}
            & \hspace{-0.07in}   2 & \hspace{-0.1in} 0.76 \hspace{-0.07in}
            & \hspace{-0.07in} 214 & \hspace{-0.1in} 2.94 \hspace{-0.07in}
    \end{tabular}
    \caption{From top to bottom:  the sizes and $1$-norms of the
      persistence and image persistence diagrams of the first blue medusa.
      Boldface numbers correspond to dots in Figure \ref{fig:Blue}.}
    \label{tbl:Blue}
\end{table}

%%\clearpage
%%%%%%%%%%%%%%%%%%%%%%%%%%%%%%%%%%%%%%%%%%%%%%%%%%%%%%%%%%%%%%%%%%%%%%%%%%
%%%%%%%%%%%%%%%%%%%%%%%%%%%%%%%%%%%%%%%%%%%%%%%%%%%%%%%%%%%%%%%%%%%%%%%%%%
\section{Discussion}
\label{sec6}
%%%%%%%%%%%%%%%%%%%%%%%%%%%%%%%%%%%%%%%%%%%%%%%%%%%%%%%%%%%%%%%%%%%%%%%%%%
%%%%%%%%%%%%%%%%%%%%%%%%%%%%%%%%%%%%%%%%%%%%%%%%%%%%%%%%%%%%%%%%%%%%%%%%%%

This paper introduces the persistent homology analysis of a medusa
as a novel method to measure cell segregation.
To provide a proof of concept, we have computed these measurements
for simulated time series of 3D data.
The application of the method to imaged cell segregation
in zebrafish embryos is forthcoming.

The medusa introduced in this paper is related to the vineyard
described in \cite[Section VIII.1]{EdHa10}; see also \cite{CEM06}.
However, there are differences.
Specifically, the vineyard would be constructed for two parameters,
the restricting radius, $\alpha$, and the time, $t$.
The result is a richer structure, namely a collection of
curves in $\Rspace^2 \times [0,1]$,
which is therefore more difficult to understand.
In contrast, the medusas fix $\alpha$ to $\alpha_0$ and
in this way facilitate a more compact representation
of a subset of the vineyard.
This is appropriate for biological cells whose size does not
vary substantially with time,
and it gives topological information that is easier to interpret
and more directly relevant to the object of study.
However, we need to keep in mind that with this choice,
the persistence diagrams are not stable under the bottleneck distance.
The results are particularly sensitive to the radius,
which leaves holes if chosen too small and absorbs holes if chosen too big.

The work in this paper has motivated the extension of Alexander duality
from spaces to functions, as proved in \cite{EdKe12}.
In particular, the Euclidean Shore Theorem in that paper
states that the persistence diagram of the time function
restricted to the boundary of a Voronoi medusa can be obtained from
the diagram of the time function on the Voronoi medusa.
Generically, the boundary is a $3$-manifold without boundary,
so that the time function can be understood in Morse theoretic terms,
which is sometimes convenient.

In conclusion, we note that the framework introduced in this paper
applies to general point processes that unfold in time.
The latter model a variety of problems of which we mention a few:
molecules of two fluids mixing after a shock-wave;
microbes forming microfilms;
a flock of birds getting into formation;
two teams competing in a soccer game;
galaxies moving under the influence of gravity.
The measurements taken within this framework are significantly coarser
than what could be said by studying braids \cite{Mag73} or the
loop spaces of configuration spaces needed to detect topological
differences for particles moving in $3$ dimensions \cite{CoGi02}.
This is not necessarily a disadvantage since the coarser information
is easier to compute as well as to comprehend.
%% It may be hoped that the coarser information can help identify
%% physical situations in which the refined information provided
%% by braids and look spaces is relevant.

%%\subsection*{Acknowledgments}
%%The authors thank Viktoriia Sharmanska for discussions and help with
%%$2$-dimensional computations.

%\newpage
%%%%%%%%%%%%%%%%%%%%%%%%%%%

\end{document}